\documentclass[12pt]{article}
\newcommand{\D}{\mbox{D}}
\newcommand{\barF}{\bar{F}}
\newcommand{\id}{\mbox{1$\!\!$I}}
\newcommand{\be}{\begin{equation}}
\newcommand{\ee}{\end{equation}}
\newcommand{\bea}{\begin{eqnarray}}
\newcommand{\eea}{\end{eqnarray}}

\newcommand{\no}{\nonumber\\}

\newcommand{\unityT}{{\bf \id}_{\bf T}}
\newcommand{\unityG}{\id_{\Gamma}}
\newcommand{\ptensor}{\otimes}
\newcommand{\bareta}{\bar{\eta}}
\newcommand{\barlambda}{\bar{\lambda}}
\newcommand{\barT}{\overline{T}}

\newcommand{\munu}{(\mu \leftrightarrow \nu)}
\newcommand{\albe}{(\alpha \leftrightarrow \beta)}
\begin{document}

\begin{center}
\title{Extended Gauge Theories in Euclidean Space
\\ with Higher Spin Fields}
\author{ E. Gabrielli}
\date{}
\par \maketitle
{\small {\it Departamento de F\'\i sica Te\'orica, C-XI,
Universidad Aut\'onoma de Madrid\\ 
E-28049 Madrid, Spain\\
Instituto de F\'\i sica Te\'orica, C-XVI,
Universidad Aut\'onoma de Madrid\\
E-28049 Madrid, Spain}}
\footnote{ 
E-mail: emidio.gabrielli@cern.ch}
\end{center}
\begin{abstract}
\noindent 
The extended Yang-Mills gauge theory in Euclidean space is a 
renormalizable (by power counting) gauge theory
describing a local interacting theory of scalar, vector, and 
tensor gauge fields (with maximum spin 2). 
In this article we study the quantum
aspects and various generalizations of this model in Euclidean space.
In particular the quantization of the pure gauge model in a 
common class of covariant gauges is performed. 
We generalize the pure gauge sector by including matter fermions 
in the adjoint representation of the gauge group and analyze
its N=1 and N=2 supersymmetric extensions.
We show that the maximum half-integer spin contained in these 
fermion fields in dimension 4 is 3/2.
Moreover we develop an extension of this theory so as to 
include internal gauge symmetries and the coupling to bosonic matter fields.
The spontaneous symmetry breaking of the extended gauge symmetry is also 
analyzed.
\end{abstract}
\newpage
\section{Introduction}
The interacting theories of higher spin fields \cite{vas}
have attracted a great attention mostly because of the relevant role
played by the spin-2 field graviton which, it is well known, should 
couple to any kind of particle. 
There is no doubt about the existence of higher spin composite 
particles in nature, a classical example is given by the 
observed hadronic resonances.
However, up to now, no elementary particle with spin higher than 1 has
been observed (among these the graviton).

The main theoretical problems which affect the construction of a
consistent interacting theory of elementary 
higher spin fields \cite{inconsist} in Minkowski space
can be briefly summarized
as follows: one must require the cancellation of all
negative-norm states, a cancellation which is
performed by means of higher-spin gauge invariances.
These local invariances, though, 
impose too many restrictions (on the interacting terms) 
which cannot be satisfied in many circumstances.
These restrictions
can be circumvented by relaxing some basic requirements of 
quantum field theory.
Indeed a general class of consistent interacting 
higher-spin gauge theories in dimension 4,3, and 2 exists
and describes infinitely many fields containing all the spins
\cite{highspin}.
In the formulation of these theories, though,
in order to implement the gauge symmetries, necessary to eliminate
all the negative norm states, infinitely 
many auxiliary fields must be introduced.
This mechanism induces higher derivatives
in the interaction terms and these in turn give rise to
non-locality \cite{vas}. These 
theories are of interest, however, since they also
establish a connection with string models, 
even though in the latter all the elementary higher spin 
excitations beyond the graviton are massive \cite{string}.

Up to now in Minkowski space,
the only consistent local interacting field theories 
of massless spin higher than 1/2 are the usual abelian and 
non-abelian Yang-Mills (YM) gauge theories for the spin-1 \cite{YM}, the 
gravity for the spin-2 \cite{spin2}, and the supergravity theories
\cite{sugra} for both the spin-2 and spin-3/2.
In addition if one requires these theories to be also renormalizable
then the above list would further shorten since it does not contain
the gravitational interactions.
At present string 
theories, where gravity is consistently coupled to matter and gauge 
fields of any spin \cite{string}, are largely believed to play a central role
in the solution of all these problems.

In this paper we analyze a gauge theory of higher 
spin interactions in a flat Euclidean space-time.
In this space, in fact, we shall see that it is possible to 
construct a general class of renormalizable higher spin gauge theories.
This theory was first introduced
a number of years ago in \cite{eg} where an extension to the abelian gauge
theory with scalar, vector, and tensor gauge fields was proposed
in Euclidean space. This model is described (in a 4--dimensional 
flat space-time) 
by a non-abelian $U(4)$ gauge theory of the YM's type 
where the connection field takes values in 
the usual Clifford algebra of spinors. Here
the scalar, vector, and tensor fields can be identified
as the components of the gauge connection along the Clifford 
algebra basis. 
In four dimensions the content of the maximum spin of the
gauge multiplet is a spin-2 and the full interacting lagrangian is 
renormalizable by power counting. This model, in the full basis
of Clifford algebra, contains
three spin-2 fields: two of them are in a standard representation
of the rotation group and are described by a symmetric traceless 
tensor of rank two, whereas the remaining one is 
contained in one of the irreducible representations of a tensor of
rank-3 with two antisymmetric indices \cite{eg}-\cite{mk2}.

An interesting aspect of this model
is that the gauge transformations mix, in a consistent way, different 
irreducible representations of the space-time rotation group.
Here the gauge spin-2 fields do not have the usual coupling 
to the energy momentum tensor while they do couple to the lowest spin particles
in a consistent way. 
\footnote{We will see that the standard spin-2 fields appearing in this model
cannot be identified with standard graviton field, 
at least in the weak coupling limit.}
Moreover the free gauge transformations of the standard spin-2 fields 
coincide with the usual spin-2 gauge transformations of the Fierz-Pauli field
\cite{FP}. 
\\
A controversial question is the analytical continuation 
of this theory to the Minkowski space. 
We next recall some problems related to this issue that are still open.
Since the elements of the gauge group are not invariant under
the Lorentz transformations, the question whether or not the model in 
\cite{eg} is forbidden by the Coleman--Mandula no--go theorem \cite{coleman}
might arise.
As pointed out in Ref.\cite{mk2}, the model in \cite{eg} circumvents 
the hypothesis of \cite{coleman}.
The main reason for this is that the theorem in \cite{coleman} applies only 
to the global symmetries of the $S$ matrix and
does not deal with the local symmetries of the Lagrangian 
(see Ref.\cite{sohnius} for a detailed discussion on this issue).
Moreover, since the present theory is of the YM's type, we should
expect the confinement phenomenon to arise.
If so, then the physical spectrum
will be described by gauge invariant operators, such as for 
example the hadron states or glueballs in QCD, and this symmetry will not 
be manifest in the $S$ matrix of the physical states. 
However we stress that, in this model, one of the main statements of the
Coleman--Mandula theorem, which is the analytical behavior of the
$S$ matrix, can be directly checked in perturbation theory. 
In particular one can verify (by means of the analogy with the 
gluon scatterings) that, in the pure Yang--Mills sector,
the tree--level gauge--invariant amplitudes
satisfy all the analytical requirements \cite{new}.
\\
This theory is well formulated in an Euclidean space
where the gauge group is compact and, being a YM's type
gauge theory, it should satisfy also the Osterwalder-Schrader axioms
\cite{OS}.
Therefore it is possible to quantize it by means of the
standard path integral method applied to gauge theories.
However, when this theory is formulated directly into a Minkowski space
(where the gauge group is non--compact), problems with unitarity 
of the $S$ matrix might arise because of unwanted ghost states 
\cite{eg},\cite{mk2}.
Nevertheless we stress that the appearance of an indefinite metrics in the
Hilbert space, due the non--compactness of the symmetry group, 
is not always an obstacle for building a consistent theory
\cite{leewick}--\cite{ha}.
A pioneering study in this direction was started by Lee and Wick in 
Ref.\cite{leewick}.
Moreover in the literature various non--compact sigma-models with indefinite
metrics also appear in some extended supergravities when the reduction 
to four dimension is considered \cite{noncompactsigma}.
The general conclusion for these models 
is that a unitary $S$ matrix in the physical subspace 
can be obtained from a pseudo-unitary S-matrix in the full Hilbert space
\cite{noncompactsigma}, \cite{noncompact}.\\
In the present model the analysis of unitarity is a more complicated issue
than in the non--compact sigma--models,
mainly because the symmetry is local and it is not an internal one.
A careful analysis of the unitarity of the $S$ matrix 
(in Minkowski space) is still missing for this theory 
and it would be worth investigating how
the unphysical ghost sector could decouple from the physical amplitudes.
The clarification of this problem
could be helpful for understanding the relation between unitarity and 
renormalizability of the spin-2 field interactions in Minkowski space.
However in the present paper we do not tackle the issue of unitarity 
and restrict ourselves to the Euclidean space where the gauge group 
is compact and the theory is consistent.

Recently in Ref.\cite{mk2} an interesting proposal to include fermions in
the model in\cite{eg} has been given.
Whereas in Ref.\cite{mk2} only the sub-group $SO(4)$ is considered, in this
article we shall see
that it is straightforward to extend these fermion 
couplings to the larger group
$U(4)$ which was first considered in Ref.\cite{eg}.
Moreover in this work we generalize the model in \cite{eg} 
to its N=1 and N=2 Euclidean supersymmetric extensions.

The paper is organized as follows.
In section [2] we briefly recall the model proposed in Ref.\cite{eg}
and analyze the free particle spectrum.
In section [3] we quantize this model in a covariant gauge 
and give the expression for the ghost lagrangian.
In section [4], by following the approach of Ref.\cite{mk2},
we generalize the model in \cite{eg} by including fermions in the 
adjoint representation of the gauge group 
and analyze its supersymmetric extensions.
In section [5] we extend the model in \cite{eg} so as 
to include the standard internal gauge symmetries.
The expression for the unified gauge lagrangian,
which includes the internal $SU(N)$ gauge group,
is given,
together with the corresponding infinitesimal gauge transformations,
in appendix.
In section [6] we study the coupling of the 
extended gauge fields with bosonic matter fields 
and make some remarks on possible couplings
with ordinary matter and gauge fields.
In section [7], the 
spontaneous symmetry breaking of the extended gauge symmetry is analyzed.
Finally the last section is devoted to our conclusions.
\section{Pure Gauge Action}

In the model proposed in Ref.\cite{eg} the usual
abelian gauge transformations have been extended to non-abelian ones which
mix fields of different integer spin.
As a consequence the elements of the gauge
group transform non-trivially under coordinate rotations.
In addition the lagrangian, which is invariant under these extended
gauge transformations, describes a 
local interacting gauge theory of higher spin fields.
This model has the attractive feature that it is 
renormalizable by power counting. (Due to the gauge invariance, 
we believe that the model is also
fully renormalizable, however we do not tackle 
this issue in the present article.)
Moreover another interesting characteristic is that the maximum value of the spin
contents ($S$) of the gauge multiplet is fixed by the space-time dimension $d$;
in particular for $d=4$ we have $S=2$.

Before presenting our analysis we briefly recall the model proposed
in Ref.\cite{eg}. One first considers a
spinorial-vectorial field 
$\hat{A}^{ij}_{\mu}(x)$ in Euclidean four dimensional space, where 
$i$ and $j$ are indices which belong to the Dirac spinorial space
$(i,j=1,\dots, 4)$.
In particular this field is defined to transform under the Euclidean 
coordinate rotation\footnote{
We use the convention
to sum up the same indices and the Euclidean
metric is given by $\delta_{\mu\nu}=\delta^{\mu}_{\nu}=diag(1,1,1,1)$.}
\be
x\to x^{\prime}_{\mu}=\Lambda_{\mu}^{~\nu} x_{\nu},~~
{\rm with}~~
\Lambda_{~\mu}^{\alpha}\Lambda_{\alpha}^{~\nu}=\delta_{\mu}^{\nu}
\label{xtransf}
\ee
as follows
\be
\hat{A}(x)_{\mu}\to\hat{A}^{\prime}_{\mu}(x')=
S(\Lambda)\hat{A}_{\nu}(x)S^{-1}(\Lambda)
\Lambda_{\mu}^{~\nu}.
\label{Axtransf}
\ee
Note that in (2) $S(\Lambda)$ is the usual 
spinorial representation of the rotation group $O(4)$ which is given by
\be
S(\Lambda)=\exp{
(-\frac{\imath}{4}\sigma_{\mu\nu}\omega^{\mu\nu})},~~{\rm with}~~
S(\Lambda)\gamma_{\mu} S^{-1}(\Lambda)=\Lambda^{\nu}_{~\mu}
\gamma_{\nu},
\label{Srepr}
\ee
where $\omega_{\mu\nu}$ is a function 
\footnote{In the case of infinitesimal transformations we have 
$\Lambda_{\mu\nu}=\delta_{\mu\nu}+
\omega_{\mu\nu} +{\cal O}(\omega^2)$, with $\omega_{\mu\nu}=-\omega_{\nu\mu}$
The exact relationship
between $\Lambda$ and $\omega$ can be found, for example, in 
Ref. \cite{gitman}.}
of $\Lambda_{\mu\nu}$,
$\sigma_{\mu\nu}=\imath[\gamma_{\mu},\gamma_{\nu}]/2$, and $\gamma_{\mu}$
are the usual Dirac matrices satisfying the relation
$\{\gamma_{\mu},\gamma_{\nu}\}=2\delta_{\mu\nu}$.
In Ref.\cite{eg} it was pointed out that it
is very useful to decompose the field $\hat{A}_{\mu}$ along
the Clifford algebra basis; this is done in the following manner
\be
\hat{A}_{\mu}(x)=A_{\mu}(x)\unityG +\bar{A}_{\mu}(x)\gamma_{5}+
T_{\mu\alpha}(x)\gamma^{\alpha}+\bar{T}_{\mu\alpha}(x)\imath
\gamma^{\alpha}\gamma_{5}
+C_{\mu\alpha\beta}(x){\sigma^{\alpha\beta}\over{\sqrt{2}}},
\label{Adecomp}
\ee
where $\unityG$ is the unity matrix in the Clifford algebra.
Indeed in this decomposition 
the indices of the fields which label the Clifford algebra
basis are vectorial indices under $O(4)$ rotations (the reader can
easily check this property by means of Eqs.(\ref{Axtransf}) and (\ref{Srepr})).
This implies that the fields $\bar{A}_{\mu}$, $T_{\mu\nu}$ (and analogously
$~\barT_{\mu\nu}$) and ~$C_{\mu\alpha\beta}$, respectively, 
transform as a vector and tensors of rank 2 and 3.\\
It is worth observing that some components of 
$\hat{A}_{\mu}$, namely $T_{\mu\alpha}$, 
$\bar{T}_{\mu\alpha}$, and $C_{\mu\alpha\beta}$, can be further
decomposed in irreducible representation of $O(4)$ as follows
\bea
T_{\mu\nu} &=& \frac{1}{4}\delta_{\mu\nu}\phi + S_{\mu\nu} + 
V^{A}_{\mu\nu}+V^{S}_{\mu\nu},\no
\overline{T}_{\mu\nu} &=& \frac{1}{4}\delta_{\mu\nu}\bar{\phi} + 
\overline{S}_{\mu\nu} + \overline{V}^A_{\mu\nu}+\overline{V}^S_{\mu\nu},\no
C_{\mu\alpha\beta}&=&\frac{1}{3}(\delta_{\mu\alpha}B_{\beta}-
\delta_{\mu\beta}B_{\alpha})
-\frac{1}{6}\epsilon_{\mu\alpha\beta\delta}\bar{B}^{\delta}+
D^S_{\mu\alpha\beta}+D^A_{\mu\alpha\beta},
\label{IRdecomp}
\eea
where $\epsilon_{\mu\nu\alpha\beta}$ is the
complete antisymmetric tensor. We now give some clarifications about
the fields appearing in Eq.(\ref{IRdecomp}):
$\phi$ and $\bar{\phi}$ are scalars, $B_{\beta}$ and 
$\bar{B}_{\beta}$ are vectors
while $S^{\mu\nu}$ is a symmetric traceless tensor of rank-2.
The fields $V^{S}_{\mu\nu}$ and  $V^{A}_{\mu\nu}$
(and analogously $\overline{V}^{S,A}_{\mu\nu}$) 
are antisymmetric tensors in the (1,0) and (0,1) representations, respectively.
(Note that with the notation (x,y) we refer to the usual $SU(2)\times SU(2)$
complex spinorial representation of the rotation group $O(4)$ \cite{ramond}.)
The tensor fields $D^{S}_{\mu\alpha\beta}$ and $D^{A}_{\mu\alpha\beta}$
belong to the $(3/2,1)$ and $(1,3/2)$ representations respectively; they
are antisymmetric in the $\alpha,~\beta$ indices and are traceless.
Moreover $V^{(S,A)}_{\alpha\beta}$ and 
$D^{(S,A)}_{\mu\alpha\beta}$ satisfy the following 
self-duality conditions \cite{mk2}
\bea
V^{(S,A)}_{\mu\nu}&=&\pm\frac{1}{2}\epsilon_{\mu\nu}^{~~\alpha\beta} 
V^{(S,A)}_{\alpha\beta},\no
D^{(S,A)}_{\mu\alpha\beta}&=&\pm\frac{1}{2}
\epsilon_{\alpha\beta}^{~~\gamma\delta}
D^{(S,A)}_{\mu\gamma\delta},
\label{selfcond}
\eea
where the signs $(+)$ and $(-)$ refer to $(S)$ and $(A)$ 
respectively.\footnote{We recall 
that in Minkowski space it is not possible to impose
self-dual (or antiself-dual) conditions on real fields.
The self- (or antiself) 
duality conditions can be imposed only on complex fields.
Therefore in Minkowski space each combination $V^S_{\mu\nu}+V^A_{\mu\nu}$ 
(and analogously the $\overline{V}^{(S,A)}_{\mu\nu}$ fields) and 
$D^S_{\mu\alpha\beta}+D^A_{\mu\alpha\beta}$, 
which appear in Eq.(\ref{IRdecomp}),
are replaced by only one irreducible field representation.}
\\
We now see that, by using the self-duality conditions (\ref{selfcond}),
the $D^{(S,A)}_{\mu\alpha\beta}$ and 
$V^{(S,A)}_{\alpha\beta}$ tensor fields have
8 and 3 degrees of freedom, respectively.
Note that, if the fields are massive\cite{weinbth},
a spin-2 field is contained in each of the tensors
$S_{\mu\nu},~\bar{S}_{\mu\nu}$ and $D^{(S,A)}_{\mu\alpha\beta}$. 
On the contrary, if the spin-2 fields are massless,
according to the Weinberg theorem\cite{weinbth}
only the left-handed and right-handed polarizations 
are consistently described by the $D^{S}_{\mu\alpha\beta}$ and
$D^{A}_{\mu\alpha\beta}$ fields, respectively. 
Therefore, if parity is conserved, one may conclude that a
massless spin-2 can be described by the reducible field 
$D_{\mu\alpha\beta}=D^{S}_{\mu\alpha\beta}+ D^{A}_{\mu\alpha\beta}$.

Returning to the model in \cite{eg}, 
the succeeding step is to promote the field
$\hat{A}_{\mu}$ to a gauge connection by requiring that
$\hat{A}_{\mu}$ transforms under a local gauge transformation $U(x)$ 
as follows
\be
\hat{A}^{G}_{\mu}(x)=
U(x)\hat{A}_{\mu}(x)U^{-1}(x)+{1\over{\imath g}}U(x)\partial_{\mu} 
U^{-1}(x),
\label{Agtransf}
\ee
where $U(x)$, which belongs to U(4),
acts on the spinorial indices of $\hat{A}_{\mu}$.
In \cite{eg} it is required that the new gauge field $\hat{A}_{\mu}^{G}$ 
should transform, under coordinate rotations, as $\hat{A}_{\mu}$ 
in Eq.(\ref{Axtransf}). Because of this requirement
the transformation (under coordinate rotations) properties of $U(x)$
\be
U'(x')=S(\Lambda)U(x)S^{-1}(\Lambda)
\label{Uxtransf}
\ee
also follow.
\\ 
Now by means of the exponential representation one can express $U(x)$
as follows
\be
U(x)=\exp{\left\{\imath \hat{\epsilon}(x)\right\}},
\label{Udecomp}
\ee
where 
\be
\hat{\epsilon}(x)=\epsilon(x)\unityG+\bar{\epsilon}(x)\gamma_{5}+
\epsilon_{\mu}(x)\gamma^{\mu}+\bar{\epsilon}_{\mu}(x)
\imath \gamma^{\mu}\gamma_{5}
+\epsilon_{\mu\nu}(x){\sigma^{\mu\nu}\over{\sqrt{2}}}.
\label{eps}
\ee
Note that the indices which label the basis in Eq.(\ref{eps}),
due to the transformation in (\ref{Uxtransf}) and Eq.(\ref{Srepr}),
transform as vectorial indices under coordinate rotations.
Finally the compact expression of the lagrangian $L_E$,
which is invariant under the gauge 
transformations (\ref{Agtransf}), is given by
\bea
L_{E}(x)&=&{1\over{16}}Tr[\hat{F}_{\mu\nu}\hat{F}^{\mu\nu}],\no
\hat{F}_{\mu\nu}(x)&=&\partial_{\mu}\hat{A}_{\nu}
-\partial_{\nu}\hat{A}_{\mu}+\imath g[\hat{A}_{\mu},\hat{A}_{\nu}],
\label{purelag}
\eea
where in the above expression the commutator and the trace are taken on the
Clifford algebra. 
By using the component fields given in
(\ref{Adecomp}), the lagrangian in (\ref{purelag}) takes the following form
\be
L_{E}=L_{0}+gL_{1}+g^2L_{2},
\label{Lpure}
\ee
where
\bea
L_{0}&=& \frac{1}{2}\left\{
(\partial_{\alpha}T_{\beta\gamma}\partial^{\alpha}T^{\beta\gamma}
-\partial_{\alpha}T_{\beta\gamma}\partial^{\beta}T^{\alpha\gamma})+
(\partial_{\alpha}\bar{T}_{\beta\gamma}\partial^{\alpha}
\bar{T}^{\beta\gamma}-
\partial_{\alpha}
\bar{T}_{\beta\gamma}\partial^{\beta}\bar{T}^{\alpha\gamma})\right.\no
&+&\left.(\partial_{\alpha}\bar{A}_{\beta}\partial^{\alpha}\bar{A}^{\beta}
-\partial_{\alpha}\bar{A}_{\beta}\partial^{\beta}\bar{A}^{\alpha})+
(\partial_{\alpha}A_{\beta}\partial^{\alpha}A^{\beta}
-\partial_{\alpha}A_{\beta}\partial^{\beta}A^{\alpha})\right.\no
&+&\left.(\partial_{\alpha}C_{\beta\gamma\delta}
\partial^{\alpha}C^{\beta\gamma\delta}
-\partial_{\alpha}C_{\beta\gamma\delta}\partial^{\beta}
C^{\alpha\gamma\delta})\right\},
\label{Lfree}
\eea
\bea
L_{1}&=& 2\left\{\sqrt{2}[
T_{\alpha\beta}C^{\gamma\beta\delta}(\partial_{\gamma}T^{\alpha}_{~~\delta}
-\partial^{\alpha}T_{\gamma\delta})
+\bar{T}_{\alpha\beta}C^{\gamma\beta\delta}(\partial_{\gamma}
\bar{T}^{\alpha}_{~~\delta}
-\partial^{\alpha}\bar{T}_{\gamma\delta})\right.\no
&+&\left.\bar{A}^{\mu}[
T^{\alpha\beta}
(\partial_{\alpha}\bar{T}_{\mu\beta}
-\partial_{\mu}\bar{T}_{\alpha\beta})+
\bar{T}^{\alpha\beta}(\partial_{\mu}T_{\alpha\beta}
-\partial_{\alpha}T_{\mu\beta})]\right.\no
&+&\left.[\sqrt{2}C_{\alpha\beta\gamma}C^{\delta\gamma\mu}
-{1\over{\sqrt{2}}}(T_{\alpha\beta}T^{\delta\mu}
+\bar{T}_{\alpha\beta}\bar{T}^{\delta\mu})][
\partial_{\delta}C^{\alpha\beta}_{~~~\mu}-
\partial^{\alpha}C_{\delta~~\mu}^{~\beta}]\right.\no
&+&\left.T_{\alpha\beta}\bar{T}^{\gamma\beta}
(\partial_{\gamma}\bar{A}^{\alpha}-\partial^{\alpha}\bar{A}_{\gamma})
\right\},
\label{Lint1}
\eea
\noindent and
\bea
L_{2}&=& 4[ C_{\alpha\beta\gamma}C^{\alpha\beta\delta}
 C^{\mu\gamma\nu}C_{\mu\delta\nu}
 -C_{\alpha~~\gamma}^{~\beta} C^{\alpha\delta\mu}
C_{\nu\beta\delta}C^{\nu\gamma}_{~~~\mu}]\no
&+&T_{\alpha\beta}T^{\alpha\beta}T_{\gamma\delta}T^{\gamma\delta}
-T_{\alpha\beta}T^{\alpha\gamma}T_{\delta\gamma}T^{\delta\beta}
+\bar{T}_{\alpha\beta}\bar{T}^{\alpha\beta}\bar{T}_{\gamma\delta}
\bar{T}^{\gamma\delta}
-\bar{T}_{\alpha\beta}\bar{T}^{\alpha\gamma}
\bar{T}_{\delta\gamma}\bar{T}^{\delta\beta}\no
&+&
4[T^{\alpha\beta}T_{\alpha\gamma}C^{\delta\gamma\mu}C_{\delta\beta\mu}+
T^{\alpha\beta}T_{\gamma\delta}C^{\gamma\delta\mu}C_{\alpha\beta\mu}
-2T_{\alpha\beta}T_{\gamma\delta}C^{\alpha\delta\mu}C^{\gamma\beta}_{~~~\mu}]
\no
&+&4[\bar{T}^{\alpha\beta}\bar{T}_{\alpha\gamma}
C^{\delta\gamma\mu}C_{\delta\beta\mu}+
\bar{T}^{\alpha\beta}\bar{T}_{\gamma\delta}
C^{\gamma\delta\mu}C_{\alpha\beta\mu}
-2\bar{T}_{\alpha\beta}\bar{T}_{\gamma\delta}
C^{\alpha\delta\mu}C^{\gamma\beta}_{~~~\mu}]\no
&+&2[T_{\alpha\beta}T^{\alpha\gamma}\bar{T}^{\delta\beta}
\bar{T}_{\delta\gamma}
+T_{\alpha\beta}\bar{T}^{\alpha\beta}T_{\gamma\delta}\bar{T}^{\gamma\delta}
-2T_{\alpha\beta}T_{\gamma\delta}\bar{T}^{\alpha\delta}
\bar{T}^{\gamma\beta}]\no
&+&2[T_{\alpha\beta}T^{\alpha\beta}\bar{A}_{\mu}\bar{A}^{\mu}
 -T_{\beta\alpha}T^{\delta\alpha}\bar{A}^{\beta}\bar{A}_{\delta}
 +\bar{T}_{\alpha\beta}\bar{T}^{\alpha\beta}\bar{A}_{\mu}\bar{A}^{\mu}
 -\bar{T}_{\beta\alpha}\bar{T}^{\delta\alpha}\bar{A}^{\beta}
\bar{A}_{\delta}]\no
&+&4\sqrt{2}
[2T_{\alpha\beta}\bar{T}^{\alpha\gamma}C^{\mu\beta}_{~~~\gamma}\bar{A}_{\mu}
 -T_{\alpha\beta}\bar{T}_{\mu\gamma}\bar{A}^{\mu}C^{\alpha\beta\gamma}
 -T_{\mu\alpha}\bar{A}^{\mu}\bar{T}_{\beta\gamma}C^{\beta\alpha\gamma}].
\label{Lint2}
\eea
The lagrangian in (\ref{purelag})
is invariant under the following local infinitesimal transformations 
\bea
\delta A_{\mu}&=&-{1\over{g}}\partial_{\mu}\epsilon,\no
\delta \bar{A}_{\mu}&=&-{1\over{g}}\partial_{\mu}\bar{\epsilon}+
2(\bar{\epsilon}^{\nu}T_{\mu\nu}-\epsilon^{\nu}\bar{T}_{\mu\nu}),\no
\delta T_{\mu\nu}&=&-{1\over{g}}\partial_{\mu}\epsilon_{\nu}
+2(\bar{\epsilon}\bar{T}_{\mu\nu}+\sqrt{2}\epsilon^{\alpha}C_{\mu\nu\alpha}-
\bar{\epsilon}_{\nu}\bar{A}_{\mu}+\sqrt{2}T_{\mu\alpha}\epsilon^{\alpha}_{~\nu}),
\no
\delta \bar{T}_{\mu\nu}&=&-{1\over{g}}\partial_{\mu}\bar{\epsilon}_{\nu}
+2(-\bar{\epsilon}T_{\mu\nu}+\sqrt{2}\bar{\epsilon}^{\alpha}C_{\mu\nu\alpha}+
\epsilon_{\nu}\bar{A}_{\mu}+\sqrt{2}\bar{T}_{\mu\alpha}\epsilon^{\alpha}_{~\nu}),
\no
\delta C_{\mu\alpha\nu}&=&-{1\over{g}}\partial_{\mu}\epsilon_{\alpha\nu}+
\sqrt{2}(\epsilon_{\alpha}T_{\mu\nu}-\epsilon_{\nu}T_{\mu\alpha})+
\sqrt{2}(\bar{\epsilon}_{\alpha}\bar{T}_{\mu\nu}-
\bar{\epsilon}_{\nu}\bar{T}_{\mu\alpha})+
\no
&&2\sqrt{2}(\epsilon^{\beta}_{~\nu}C_{\mu\alpha\beta}-
\epsilon^{\beta}_{~\alpha}C_{\mu\nu\beta}).
\label{gaugetransf}
\eea
Clearly the $A_{\mu}$ is a free field since it corresponds to the
$U(1)$ gauge connection of $U(4)$ and the interacting theory is 
described by the $SU(4)$ gauge group.  
Note that the smallest sub-algebra of $SU(4)$ is given by the 
$\sigma_{\mu\nu}$ generators which belong to
the algebra of $SO(4)$. The smallest
gauge lagrangian, which is invariant under the $SO(4)$ gauge transformations,
is given by the terms in (\ref{Lfree}-\ref{Lint2}) 
containing only the $C_{\mu\alpha\beta}$ field. The latter
was considered in Ref.\cite{mk2}.

One interesting aspect of this model is that 
the lagrangian (\ref{Lpure}) can be written in terms 
of the $O(4)$ irreducible representations 
by inserting the fields decomposition (\ref{IRdecomp}) in the
expressions (\ref{Lfree})-(\ref{Lint2}) (see \cite{eg}).
Then the corresponding infinitesimal gauge transformations for the
irreducible fields are obtained by means of the
standard decomposition method, as previously shown
in Eq.(\ref{gaugetransf}).
For example, the corresponding infinitesimal gauge transformations for
$\delta \phi,~\delta S_{\mu\nu},~\delta V^{(S,A)}_{\mu\nu}$ are given by
\bea
\delta \phi &=& \delta T_{\mu}^{~\mu},
~\delta S_{\mu\nu}=
\frac{1}{2}\left(\delta T_{\mu\nu}+\delta T_{\nu\mu}\right)-
\frac{1}{4}\delta_{\mu\nu} \delta T_{\alpha}^{~\alpha},\no
\delta V^{(S,A)}_{\mu\nu}&=&\frac{1}{4}
\left(\delta T_{\mu\nu}-\delta T_{\nu\mu}
\pm\frac{1}{2}\delta T^{\alpha\beta} \epsilon_{\alpha\beta\mu\nu} \right),
\eea
with the obvious generalization for the other fields.
The expression for the lagrangian containing only 
the $C_{\mu\alpha\beta}$ field, in terms of the irreducible 
representation, can be found in Ref.\cite{mk2}.

We here analyze the physical degrees of freedom and the free particle spectrum
of the model introduced in \cite{eg}.
In order to do so we restrict our
analysis by only considering the free lagrangian $L_0$ 
which is invariant under the abelian sector\footnote{By abelian 
sector of the gauge transformations we mean the non-homogeneous terms in 
Eq.(\ref{gaugetransf}) containing only the derivatives.} 
of the gauge transformations.
We first consider\footnote{
In this analysis we do not consider
the free particle spectrum described by the
fields $A_{\mu}$ and $\bar{A}_{\mu}$ since it is clear from
(\ref{Lfree}) that they describe two massless spin-1 fields.}
the free lagrangian $L_0(T)$ in Eq.(\ref{Lfree})
which contains only the tensor field $T_{\mu\nu}$.
From Eq.(\ref{gaugetransf}) 
we see that it is possible to make a gauge transformation which ensures
\be
\partial^{\mu} S_{\mu\nu}=0~~{\rm and}~~\phi=0.
\label{constr1}
\ee
This result is justified in the following manner.
The free gauge transformations for 
the field $\tilde{S}_{\mu\nu}=S_{\mu\nu}+1/4\delta_{\mu\nu}\phi$
are given by $\delta \tilde{S}_{\mu\nu}=\partial_{\mu} \epsilon_{\nu}+
\partial_{\mu} \epsilon_{\nu}$. Then, by means of these gauge transformations,
the first constraint in (\ref{constr1}) can be imposed.
This  constraint,
when the on-shell massless equations for $S_{\mu\nu}$ and $\phi$ are used, 
is invariant under a new gauge transformation.
This new gauge degree of freedom allows us to eliminate the component of the 
on-shell massless scalar field  $\phi$. 
(Note that the gauge transformations for the field $\tilde{S}_{\mu\nu}$
and the gauge fixing in (\ref{constr1}) are the same ones encountered
in the spin-2 field of the Fierz-Pauli lagrangian.)
By adopting the gauge fixing (\ref{constr1}) all the gauge degrees of freedom
are used and no new constraint on $T_{\mu\nu}$ can be imposed.
In particular no constraint may be introduced for
the antisymmetric tensor field $T^{A}_{\mu\nu}=V^S_{\mu\nu}+
V^A_{\mu\nu}$.\\
Now if the fields $V^{(S,A)}_{\mu\nu}$, in the (1,0) and (0,1) representation
fields, were massive then they would describe two spin-1 fields.
However since they are massless fields, according to the Weinberg theorem
\cite{weinbth}, only one massless
spin-1 field can be associated to the $(1,0)\oplus(0,1)$ representation, 
where the right- and left-handed polarization are contained in the 
(1,0) and (0,1) representation, respectively.
Here the two residual degrees of freedom are contained in the longitudinal
components. These degrees are associated
to the transverse polarizations of the 
vectorial field $L_{\mu}\equiv \partial_{\nu} T^{\mu\nu}$.
Indeed these polarizations cannot be gauged out in this model.
The only components of the reducible tensor $T_{\mu\nu}$ which can be gauged
out correspond to the vector $\partial_{\mu} T^{\mu\nu}$.

As a consequence the physical polarizations of 
the field $T_{\mu\nu}$, in the momentum space $k$, are
the ``transverse'' ones $\epsilon^{T~(S)}_{\mu\alpha}(k)$ and
$\epsilon^{T~(A)}_{\mu\alpha}(k)$
(where $(S)$ and $(A)$ refer to the symmetric and antisymmetric
tensor in $\mu$ and $\alpha$, respectively)
and ``longitudinal'' ones $\epsilon^{L}_{\mu\alpha}(k)$
which satisfy the following conditions
\be
k^{\mu}\epsilon^{T~(A)}_{\mu\alpha}(k)=0,~~~
k^{\mu}\epsilon^{T~(S)}_{\mu\alpha}(k)=0,~~~
k^{\mu}\epsilon^{L}_{\mu\alpha}(k)=0,
\label{Tpol}
\ee
note that in the above relations $\epsilon^{L}_{\mu\alpha}$ 
has not a definite symmetry in $\mu,\alpha$ and
$k^{\alpha}\epsilon^{L}_{\mu\alpha}(k)\ne 0$. 
From Eq.(\ref{Tpol}) we have
$\epsilon^{T~(S)}_{\mu\alpha}(k)$ and $\epsilon^{T~(A)}_{\mu\alpha}(k)$
each contain only two independent polarizations. 
On the contrary  in $\epsilon^{L}_{\mu\alpha}$
there are four independent polarizations, each of which 
correspond to one massless spin-1 and two massless spin-0.
As a result the on-shell $T_{\mu\nu}$ field 
describes the following spectrum:
one massless spin-2, two massless spin-1, and two massless spin-0.
Therefore in total we count $3\times {\bf 2}~+~2\times {\bf 1}={\bf 8}$
degrees of freedom for the on-shell 
$T_{\mu\nu}$ field; this result is in agreement with the naive counting
based on the gauge degrees of freedom.

We next analyze the particle spectrum described by the free lagrangian
$L_0(C)$ in Eq.(\ref{Lfree}) which contains only the field 
$C_{\mu\alpha\beta}$.
We can use the gauge degrees of freedom
\be
C_{\mu\alpha\beta}\to C_{\mu\alpha\beta} +\partial_{\mu} \epsilon_{\alpha\beta}
\ee
in order to set the following transversality constraints on the fields 
$D_{\mu\alpha\beta}^{(S,A)}$
\be 
\partial^{\mu} D^{(S,A)}_{\mu\alpha\beta}=0.
\label{constr2}
\ee 
The results in Ref.\cite{weinbth} enable us to see that in the massless
case the $D^{(S)}_{\mu\alpha\beta}$ 
which satisfies Eq.(\ref{constr2}) describes two degrees of freedom.
These degrees of freedom 
correspond to a right-handed spin-2 and a right-handed spin-1 field.
Analogously we can see that the $D^{(A)}_{\mu\alpha\beta}$ describes the 
corresponding left-handed ones. Therefore, since in this model 
parity is conserved, the physical polarizations of the reducible field 
$D^{(S)}_{\mu\alpha\beta}+D^{(A)}_{\mu\alpha\beta}$ will describe a 
massless spin-2 and spin-1 field.

As a result of the gauge-fixing in (\ref{constr2}) there are no
gauge degrees of freedom available
which would enable us to eliminate other components in the vector 
fields $B_{\mu}$ and $\bar{B}_{\mu}$.
Thus the field $B_{\mu}$ contains 4 independent polarizations.
The two transverse one (respect to the three-momentum) correspond to a
massless spin-one polarizations whereas the longitudinal ones are
associated with massless spin-0 fields. The spectrum for $\bar{B}_{\mu}$
is obtained analogously.
As a result the on-shell $C_{\mu\alpha\beta}$ field describes the
following massless spectrum:
one spin-2, three spin-1, and four 
spin-0 fields;
so in total we count respectively $4\times {\bf 2}~+~4\times {\bf 1}={\bf 12}$ 
degrees of freedom,
in agreement with the naive counting based on the gauge degrees of freedom.

It is clear that the on-shell particle contents of this model 
is gauge invariant and one can reach the same conclusions on the spectrum
by using different choices for the gauge fixing.
Finally we note that the spin-0  
particles (or longitudinal photons) which appear in the 
spectrum are  strictly connected to the fact that 
some longitudinal components of the tensor or vector fields 
can not be gauged out. 
\section{Covariant Quantization}
When the model in \cite{eg} is quantized in the Euclidean space the
negative norm states are absent since 
the space-time metric is the $\delta_{\mu\nu}$ and the gauge group is compact. 
Moreover, due to the compactness of the gauge 
group, the theory can be quantized by means of the
standard path integral method.
Clearly the fact that the theory is well defined in Euclidean space
it is not enough to guarantee its analytical continuation to the
Minkowski one. 
Even if the Osterwalder-Schrader (OS) axioms
are satisfied, and in particular the property of
reflection positivity \cite{OS} is verified, one cannot use here
the reconstruction theorems 
\footnote{
These theorems guarantee that the Wightman functions (which satisfy the
Wightman axioms)
can be completely reconstructed from the analytical continuation of 
the corresponding Schwinger functions, provided that these last one
satisfy the axioms in \cite{OS}.} of Ref.\cite{OS}. Indeed,
in the proof of these theorems, the gauge group is not changed by 
the analytical continuation to the Minkowski space.
On the contrary in the model in \cite{eg},
in order to maintain the Lorentz covariance, we must rotate the
gauge group $U(4)$ to the non-compact one $U(2,2)$ when the 
analytical continuation to Minkowski space is performed.
However, as we discussed in the introduction,
the presence of extra negative--norm 
states (induced by the non--compact groups) is not always an obstacle 
for building a consistent theory \cite{leewick}--\cite{ha}.
In particular, for this model, it should be interesting to see if 
a Lorentz invariant Hilbert subspace, where the theory
is unitary and the unphysical states decouple from the physical
amplitudes, exists. 
However in present paper we do not tackle the issue of 
unitarity in Minkowski space.

Now we analyze the covariant quantization of this model
in the most common class of covariant gauges.
In Euclidean space the path integral representation of the generating 
functional of the Green functions  $W[J]$ can be formally written as
\be
W[J]=\int D\hat{A}_{\mu}D\hat{\eta}^{\dag}D\hat{\eta}
\exp\left\{-\int d^4x \left(L_E+L_{GF}+\imath L_{GH}-Tr(\hat{J}_{\mu}
\hat{A}^{\mu})\right)\right\},
\ee
where $L_E$ is the full lagrangian given in Eq.(\ref{Lpure}) and 
$L_{GF}$ and $L_{GH}$ correspond to the gauge-fixing 
and the ghost lagrangian, respectively. In the last term the trace is taken on 
the Clifford algebra and $\hat{J}_{\mu}$, which can be 
decomposed as
$\hat{A}_{\mu}$ in Eq.(\ref{Adecomp}), is the source for the
gauge field $\hat{A}_{\mu}$.

In the present study we consider the general class of covariant gauges whose
gauge--fixing lagrangian is given by
\bea
L_{GF}&=&\frac{1}{8\xi}Tr\left[(\partial_{\mu}\hat{A}^{\mu})^2\right]
\label{gaugefix}
\eea
We restrict our analysis by considering
only the interacting theory given by the $SU(4)$ gauge group.
In this gauge the free propagators $P_{AB}$ 
(in momentum space $k_{\mu}$)
for the fields $\bar{A}_{\mu}$, $T_{\mu\nu}$, $\overline{T}_{\mu\nu}$ and 
$C_{\mu\alpha\beta}$ are given by
\bea
P_{(\bar{A}_{\mu}\bar{A}_{\nu})}&=&\frac{1}{k^2} 
\left(\delta_{\mu\nu}-(1-\xi)\frac{k_{\mu}k_{\nu}}{k^2}\right),\no
P_{(T_{\mu\alpha}T_{\nu\beta})}&=&\frac{1}{k^2}\delta_{\alpha\beta}\left(
\delta_{\mu\nu}-(1-\xi)\frac{k_{\mu}k_{\nu}}{k^2}\right),\no
P_{(\overline{T}_{\mu\alpha}\overline{T}_{\nu\beta})}&=&
P_{(T_{\mu\alpha}T_{\nu\beta})},\no
P_{(C_{\mu\alpha\gamma}C_{\nu\beta\delta})}&=&\frac{1}{k^2}
\frac{\left(\delta_{\alpha\gamma}\delta_{\beta\delta}-
\delta_{\beta\gamma}\delta_{\alpha\delta}\right)}{2}
\left(\delta_{\mu\nu}-(1-\xi)\frac{k_{\mu}k_{\nu}}{k^2}\right).
\label{propag}
\eea
The propagators in the basis of
the $O(4)$ irreducible representations can
be easily obtained from Eqs.(\ref{propag}) by means of the standard 
decomposition methods.
Note that the propagators in Eqs.(\ref{propag})
are diagonal in the basis of the $O(4)$  
irreducible representations only in the t'Hooft-Feynman gauge $\xi=1$.
However for practical calculations, such as the 
scattering amplitudes, it is more convenient to
work with the propagators in the basis of the reducible fields
$T_{\mu\nu},\overline{T}_{\mu\nu}$ and 
$C_{\mu\alpha\beta}$ instead of the irreducible ones.

For completeness we give the expression of the following ghost lagrangian
associated with the gauge fixing in (\ref{gaugefix})
\be
L_{GH}=L_{0}+g L_{I},
\label{Lghost1}
\ee
where
\bea
L_{0}&=&\partial^{\alpha}\bareta^{\star}\partial_{\alpha}\bareta 
+ \partial^{\alpha}\eta_{\nu}^{\star}\partial_{\alpha}\eta^{\nu}
+ \partial^{\alpha} \bareta_{\nu}^{\star} \partial_{\alpha}
\bareta^{\nu} +
\partial^{\alpha}\eta_{\mu\nu}^{\star}
\partial_{\alpha}\eta^{\mu \nu},\no
L_{I}&=&2\left[\bareta^{\star} T_{\mu\nu}\partial^{\mu}\bareta^{\nu}
- \bareta^{\star} \barT_{\mu\nu}\partial^{\mu}\eta^{\nu} -
\sqrt{2}\left(\eta_{\nu}^{\star}T_{\mu\alpha}\partial^{\mu} \eta^{\nu\alpha}
-\eta^{\star}_{\nu} 
C^{\mu\nu\alpha} \partial_{\mu} \eta_{\alpha}\right)\right.\no
&+&\left. \eta_{\nu}^{\star}\barT_{\mu\nu}\partial^{\mu} \bareta -
\eta_{\nu}^{\star}\bar{A}_{\mu}\partial^{\mu} \bareta^{\nu} -
\sqrt{2}\left( \bareta_{\nu}^{\star}
\barT_{\mu\alpha}\partial^{\mu} \eta^{\nu\alpha}
- \bareta_{\nu}^{\star}C^{\mu\nu\alpha}\partial_{\mu}\bareta_{\alpha}\right)
\right.\no
&-&\left.
\bareta_{\nu}^{\star}T^{\mu\nu}\partial_{\mu} \bareta
+\bareta_{\nu}^{\star}\bar{A}^{\mu}\partial_{\mu} \eta^{\nu}
+\sqrt{2}\left(
\eta_{\alpha\nu}^{\star}T^{\mu\nu}\partial_{\mu} \eta^{\alpha}+
\eta_{\alpha\nu}^{\star}\barT^{\mu\nu}\partial_{\mu} \bareta^{\alpha} \right.
\right.\no
&+&\left.\left.
2~\eta_{\alpha\nu}^{\star}C^{\mu\alpha\beta}\partial_{\mu} \eta^{\beta\nu}
\right)\right].
\label{Lghost2}
\eea
The ghost multiplet, which appears in Eqs.(\ref{Lghost2}),
is composed by the following fields: 
a complex scalar $\bareta$, two complex vectors 
$\eta_{\mu}$, $\bareta_{\mu}$, and a complex 
antisymmetric tensor $\eta_{\alpha\beta}$,
all of which are Grassman variables.
Note that vectorial ghost fields always appear when gauge 
spin-2 fields are present, a classical example is the quantum gravity.

It is worth noting that, although the physical spectrum should be
described in terms of fields which belong to 
the $O(4)$ irreducible representations, 
the renormalization properties of the lagrangian (\ref{Lpure}) 
can be directly analyzed by means of the 
$O(4)$ reducible field basis ($\bar{A}, T, \barT, C$) defined in 
(\ref{Adecomp}). Indeed in this basis and 
in the covariant gauge  (\ref{gaugefix}), 
the renormalizability of the lagrangian (\ref{Lpure}) is apparent.
In particular, in terms of the $O(4)$ reducible fields,
this lagrangian is equivalent to a standard YM's one with a particular
gauge group. Therefore the standard results on the 
renormalizability of the YM theories should also hold for this model.

Now we briefly discuss the gauge-invariant regularization. 
Due to the presence of the $\gamma_5$ matrix (which does not exist in
odd dimensions) in the gauge Clifford algebra basis, it is clear that 
the usual dimensional-regularization is not particularly 
suitable for this model. 
In order to solve this problem, in Ref.\cite{mk2} it is suggested to 
adopt the operator regularization scheme \cite{mksh}.
Here we want to point out that there exists 
another possible $SU(4)$ gauge invariant regularization scheme, 
which is also non--perturbative, for this theory:
this is provided by means of the corresponding Wilson action on the lattice
\cite{wilson}. 
Indeed on the lattice the Clifford algebra basis is taken in 4 dimensions
and the above mentioned problem of $\gamma_5$ does not exist.
Moreover, even though the
fermions matter fields are coupled to the gauge connection,
the lattice regularization does not spoil the $SU(4)$ 
gauge symmetry.
Indeed, as we will show later on,
the generator corresponding to the gauge transformation containing 
the $\gamma_5$ matrix is not connected to the ``standard'' 
chiral transformations and so the Wilson term, which is necessary to solve the
doubling problem, does respect the gauge symmetry.
Clearly, when fermions are added to the theory,
the Wilson term breaks the (global) ``standard'' chiral symmetry.
\section{Supersymmetric Extension}
In this section we consider the couplings of the gauge field
$\hat{A}_{\mu}$ to fermion matter fields which are 
in the adjoint representation of the gauge group.
In Ref. \cite{mk2} these couplings have been proposed,
and the smallest gauge sub-group $SO(4)$ studied. 
We generalize this approach by considering the larger group $SU(4)$.
Moreover we derive
the N=1 and N=2 supersymmetric extensions
of the pure gauge action (\ref{Lpure}).

In order to introduce these fermion couplings we follow the method
developed in Ref.\cite{mk2}. We define in Euclidean space 
the following fermion multiplet $\hat{\Psi}^{i j}_{k}$ where the
up indices $i,j$ and the down index $k$ are the usual Dirac indices.
Before giving the coordinate transformation rules of $\hat{\Psi}^{i j}_{k}$
some definitions are in order. In addition to the spinorial representation
of the $O(4)$ rotation group, namely $S(\Lambda)$, we introduce
another independent spinorial representation 
of this group that we will call $\bar{S}(\Lambda)$.
The matrix $\bar{S}(\Lambda)$,
in terms of a new Clifford algebra basis 
\be
\bar{\Gamma}_i=\id_{\bar{\Gamma}},~
\bar{\gamma}_5,~\bar{\gamma}_{\mu},~\bar{\gamma}_5,~
\imath \bar{\gamma}_{\mu}\bar{\gamma}_5,~\bar{\sigma}_{\mu\nu},
\label{Gammabasis}
\ee
(where $\id_{\bar{\Gamma}}$ is the unity matrix)
is assumed to commute with $S(\Lambda)$ and to have 
the same representation as $S(\Lambda)$ 
(see Eq.(\ref{Srepr}) ).
Note that each element of the  $\bar{\Gamma}_i$ basis is assumed to commute
with any other element of the Clifford $\Gamma_i$ basis.

Now we can define the following coordinate transformation properties 
of $\hat{\Psi}^{i j}_{k}$ in Euclidean space, these are
\bea
x\to x^{\prime}_{\mu}&=&\Lambda_{\mu}^{~\nu} x_{\nu},\no
\hat{\Psi}_{k}^{ij}(x)\to \hat{\Psi}_{k}^{\prime~ij}(x^{\prime})&=&
\left(\bar{S}(\Lambda)\right)_{km}
\left\{\left(S(\Lambda)\right)^{ia}\hat{\Psi}_m^{ab}(x)
\left(S^{-1}(\Lambda)\right)^{bj}\right\}
\eea
or in a more compact notation
\be
\hat{\Psi}(x)\to \hat{\Psi}^{\prime}(x^{\prime})=\bar{S}(\Lambda)
\left\{S(\Lambda)\hat{\Psi}(x)S^{-1}(\Lambda)\right\}
\label{Psixtransf}
\ee
and analogously for the adjoint field $\bar{\hat{\Psi}}$
\be
\bar{\hat{\Psi}}(x)\to 
\bar{\hat{\Psi}}^{\prime}(x^{\prime})=
\left\{S(\Lambda)\bar{\hat{\Psi}}(x)S^{-1}(\Lambda)\right\}
\bar{S}^{-1}(\Lambda),
\label{Psibarxtransf}
\ee
where the multiplication of $S(\Lambda)$ and $\bar{S}(\Lambda)$ 
matrices acts on the up and down Dirac indices,  respectively.
Note that the $S(\Lambda)$ matrix in Eq.(\ref{Psixtransf}) 
is the same matrix appearing in the coordinate 
transformation rule of the gauge potential in Eq.(\ref{Axtransf}).
We now decompose the field $\hat{\Psi}$ 
on the same basis $\Gamma_i$ of the $\hat{A}_{\mu}$ field as follows
\be
\hat{\Psi}_i^{j k}= \lambda_i (\gamma_5)^{jk} +  
\lambda^{\mu}_i (\gamma_{\mu})^{jk} +
 \lambda^{\mu}_{5i}~\imath(\gamma_5 \gamma_{\mu})^{jk} + 
\lambda^{\mu\nu}_i \frac{(\sigma_{\mu\nu})^{jk}}{{\scriptstyle \sqrt{2}}}.
\label{psidecomp}
\ee
In the sequel the notation for the spinorial down index ``i'' in the
component fields, appearing in (\ref{psidecomp}), will be suppressed.
As a consequence of the coordinate transformations (\ref{Psixtransf}),
the component fields  
$\lambda$, $\lambda_{\mu}$ (or analogously $\lambda_{5\mu}$), and
$\lambda_{\mu\nu}$ transform in the following manner
\bea
\lambda\to \lambda^{\prime}&=&\bar{S}(\Lambda) \lambda,\no
\lambda_{\mu}\to \lambda^{\prime}_{\mu}&=&
\Lambda^{~\alpha}_{\mu} \bar{S}(\Lambda)\lambda_{\alpha},\no
\lambda_{\mu\nu}\to \lambda^{\prime}_{\mu\nu}&=&
\frac{1}{2}\left(\Lambda^{~\alpha}_{\mu}\Lambda^{~\beta}_{\nu} 
-\Lambda^{~\beta}_{\mu}\Lambda^{~\alpha}_{\nu} \right)
\bar{S}(\Lambda) \lambda_{\alpha\beta}.
\label{Compxtransf}
\eea
There $\lambda$ (respectively $\lambda_{\mu}$ and $\lambda_{\mu\nu}$) transform
as a spin-1/2 a (respectively spin-3/2) field.

It is interesting to observe that
the fermion fields $\lambda^{\mu}_{i}$ and $\lambda^{\mu\nu}_{i}$ 
can be decomposed into irreducible representations of the 
$O(4)$ coordinate transformations as follows 
\footnote{
The spinorial indices ``i,j'' appearing in (\ref{Psidecomp})
have been temporary reintroduced to avoid confusions with the notation,
and the same indices are intended to be summed up.}
\bea
\lambda^{\mu}_{i}&=&\frac{1}{\sqrt{2}}
\bar{\gamma}^{\mu}_{ij} \psi_{j}+\psi^{\mu}_{i},\no
\lambda^{\mu}_{5i}&=&\frac{1}{\sqrt{2}} 
\bar{\gamma}^{\mu}_{ij}\psi_{5j}+\psi^{\mu}_{5i},\no
\lambda^{\mu\nu}_{i}&=& \frac{\imath}{2}
\bar{\sigma}^{\mu\nu}_{i j} \xi_{j}+\frac{1}{4}\left(
\bar{\gamma}^{\mu}_{i j}\xi^{\nu}_{j}-
\bar{\gamma}^{\nu}_{i j}\xi^{\mu}_{j}\right)+\psi^{\mu\nu}_i,
\label{Psidecomp}
\eea
where (in compact notation)
\bea
\bar{\gamma}_{\mu}\psi^{\mu}&=&~~\bar{\gamma}_{\mu}\psi_5^{\mu}=~~
\bar{\gamma}_{\mu}\xi^{\mu}=0,~~
\bar{\gamma}_{\mu}\psi^{\mu\nu}=0,\no
\psi^{\mu\nu}&=&-\psi^{\nu\mu}.
\eea
The fields $\psi$, $\psi_5$, $\xi$ ,
( resp. $\psi^{\mu}$, 
$\psi_5^{\mu}$, $\xi^{\mu}$, $\psi^{\mu\nu}$) describe
spin-1/2 (respectively spin-3/2) fields. 
By means of Eq.(\ref{Compxtransf}) it is straightforward to prove that
the decompositions (\ref{Psidecomp}) are $O(4)$ irreducible.

In order to couple the field $\hat{\Psi}$ to the gauge connection 
$\hat{A}_{\mu}$ we need to require that, under the 
gauge transformations $U(x)$, the field $\hat{\Psi}$, and
its adjoint $\bar{\hat{\Psi}}$, transform as follows
\bea
\hat{\Psi}_{k}^{ij}(x)\to \hat{\Psi}_{k}^{G ij}(x)&=&
\left(U(x)\right)^{ia}\hat{\Psi}_k^{ab}(x)\left(U^{-1}(x)\right)^{bj}\no
\bar{\hat{\Psi}}_{k}^{ij}(x)\to \bar{\hat{\Psi}}_{k}^{G ij}(x)&=&
\left(U(x)\right)^{ia}\bar{\hat{\Psi}}_k^{ab}(x)\left(U^{-1}(x)\right)^{bj}
\label{Psigtransf}
\eea
where, as usual, the sum over repeated indices is understood.
As a result the covariant derivative $\hat{D}_{\mu}$ acting on $\hat{\Psi}$
is given by
\be
\hat{D}_{\mu}\hat{\Psi}=\partial_{\mu}\hat{\Psi}-\imath 
g\left[\hat{\Psi},\hat{A}_{\mu}\right],
\label{covderiv}
\ee
where the commutator is taken on the $\Gamma_i$ Clifford algebra basis
and $g$ is the same coupling appearing in the lagrangian (\ref{Lpure}).

We now analyze the N=1 and N=2 supersymmetric extensions of the action 
(\ref{Lpure}), by 
first recalling the known technical solutions 
for constructing supersymmetric theories in Euclidean space.\\
In the four--spinor formalism, N=1 
supersymmetry (SUSY) requires the existence of 
Majorana fermions. However the Majorana reality condition 
$\psi=C~\bar{\psi}^t$ (where $C$ is the charge conjugation matrix,
$\bar{\psi}$ is the adjoint of $\psi$, 
and the suffix $t$ stands for transpose) is 
inconsistent in Euclidean space \cite{ramond}, \cite{NW}.\footnote{
This can be simply understood by means of the $SO(4)$ coordinate transformation
properties of the Weyl spinors $\lambda_L$, $\lambda_R$ in which a 
four-spinor $\lambda$ can be decomposed, namely 
$\lambda=(\lambda_L,\lambda_R)$: 
in Euclidean space the $\lambda_L$ and $\lambda_R$ transform 
independently under $SO(4)$, and so they 
can not be related by complex conjugation 
as in the relativistic theory.
This is connected to the fact that, due to the compactness of the 
Euclidean $SO(4)$ rotational group, the generators of the two $SU(2)$ groups
associated to $SO(4)$, 
are completely independent, while for the Lorentz group they are 
connected by charge conjugation. Therefore real four-spinor fields 
can not exist in Euclidean space.}
This is apparently a technical,
but not fundamental difficulty to implement supersymmetry 
in Euclidean space.\footnote{
For instance, one can always impose
simplectic Majorana 
conditions $\psi(1)=C~\bar{\psi}(2)^t$, $\psi(2)=-C~\bar{\psi}^t(1)$,
by means of two independent
four-spinors $\psi(1)$ and $\psi(2)$, where
$\bar{\psi}(1),~\bar{\psi}(2)$ are the corresponding adjoint ones.}\\
In the N=1 supersymmetric YM (SYM) theory in Euclidean space, 
by means of a special definition for Euclidean Majorana spinors 
\cite{nicolai}, \cite{FO}, supersymmetry can be restored 
at the price of giving up the hermiticity of the action.
Then, by construction, 
the expectation values (Schwinger functions),
generated by the Euclidean supersymmetric model, are the analytical 
continuations of the corresponding Green functions in Minkowski
space \cite{nicolai}.
Indeed, as observed in Ref. \cite{nicolai}, in Euclidean space the relevant
notion is not hermiticity, but rather  Osterwalder-Schrader reflection 
positivity \cite{OS} which guarantees 
the above mentioned analytical continuation.
Moreover the Euclidean N=1 SUSY transformations are not hermitean.
However this is not a problem, since
the SUSY transformations are just a formal
device in order to obtain SUSY Ward identities \cite{nicolai}.

In an other approach, proposed by Zumino \cite{zumino}, the hermiticity of
the action is retained at the price of giving up the
explicit connection between relativistic and Euclidean field theory.
In particular in the Zumino model \cite{zumino} the number of fermionic 
degrees of freedom is doubled (with respect to
N=1 SYM) in order to get an hermitean action, but
additional bosonic (scalar) matter fields should be added to the
N=1 SYM action in order to restore
the balance between fermionic and bosonic degrees of freedom. 
As a consequence the Zumino model is an 
N=2 SYM theory and therefore the analytical continuation 
with the relativistic N=1 SYM theory is lost.

We start our analysis with the N=1 SUSY extension of the lagrangian 
(\ref{Lpure}).
Paralleling the technique developed in first reference of \cite{nicolai},
the first step is to introduce, in addition to the complex field $\hat{\Psi}$ 
in Eq.(\ref{psidecomp}),
a completely independent (complex) field $\bar{\hat{\Xi}}$ 
which transforms as the adjoint of 
$\hat{\Psi}$ in Eq.(\ref{Psibarxtransf}).
The expression of $\bar{\hat{\Xi}}$  in components is given by 
\be
\bar{\hat{\Xi}}_i^{j k}= \barlambda_i (\gamma_5)^{jk} +  
\barlambda^{\mu}_i (\gamma_{\mu})^{jk} +
 \barlambda^{\mu}_{5i}~\imath(\gamma_5 \gamma_{\mu})^{jk} + 
\barlambda^{\mu\nu}_i \frac{(\sigma_{\mu\nu})^{jk}}{{\scriptstyle \sqrt{2}}}.
\label{psibardecomp}
\ee
The next step consists in imposing the 
following Majorana--like conditions on the component fields 
of $\bar{\hat{\Xi}}$ and $\hat{\Psi}$, namely
\be
\lambda \equiv C\bar{\lambda}^{t},
~~\lambda_{\mu}\equiv C\bar{\lambda}^{t}_{\mu},~~
\lambda_{5 \mu}\equiv C\bar{\lambda}^{t}_{5 \mu},~~
\lambda_{\mu\nu}\equiv C\bar{\lambda}^{t}_{\mu\nu}
\label{majorana}
\ee
where $t$ stands for transpose, $C$ (the charge conjugation matrix)
is defined as $C^{\dag}C=1$ and $C^{-1}=-C$, and the standard spinorial
multiplication between $C$ and the fermion fields $\lambda_a$ or 
$\barlambda_a$ is understood.
Note that the above relations (\ref{majorana}) can be now 
consistently satisfied since $\bar{\lambda}_a\neq \lambda^{\dag}_a\gamma_0$.
In other words, Eqs.(\ref{majorana})
are just definitions for the fields $\bar{\lambda}_a$ \cite{nicolai}.
Then, in order to implement the analytical continuation to Minkowski space,
one should require that $\lambda_a$ are Euclidean 
Majorana spinors, whose formal definition can be found
in the first reference of \cite{nicolai}.\footnote{
We recall that the Euclidean 
Majorana spinors (which are complex four-spinors) 
are {\it by definition} those operators
in Euclidean Fock space which generate the analytical continuation of 
the corresponding Wightman functions of the relativistic theory 
\cite{nicolai}.}

Finally the $O(4)$--gauge--invariant lagrangian $L_{F}$ for the 
fermion sector is given by
\be
L_{F}=\frac{1}{8}
\left\{
Tr \left(\bar{\hat{\Xi}}\bar{\gamma}_{\mu} \partial^{\mu} \hat{\Psi} \right)
-\imath g Tr\left(\bar{\hat{\Xi}}\bar{\gamma}_{\mu} 
\left[\hat{\Psi},\hat{A}^{\mu}\right]\right)
\right\},
\label{LagPsi1}
\ee
where, in component notations, we have\footnote{
Note that in Eq.(\ref{LagPsi2}) we have eliminated 
from the notation the ``bar'' over 
the $\bar{\gamma}_{\mu}$ matrices
since in the following we work only with the component
fields along the $\Gamma_i$ basis.}

\bea
L_{F}&=&L_{F}^0+g L_F^I,\no
L_F^0&=&\frac{1}{2}\left\{\bar{\lambda}\gamma^{\mu}\partial_{\mu}\lambda
+\bar{\lambda}_{\alpha}\gamma^{\mu}\partial_{\mu}\lambda^{\alpha}
+\bar{\lambda}_{5\alpha}\gamma^{\mu}\partial_{\mu}\lambda_5^{\alpha}
+\bar{\lambda}_{\alpha\beta}\gamma^{\mu}\partial_{\mu}
\lambda^{\alpha\beta}\right\},\no
L_F^I&=&2 \bar{A}^{\mu} 
\left(\barlambda_{\alpha}\gamma_{\mu}\lambda_5^{\alpha}\right)-
2~T^{\mu\alpha}\left[\left(\barlambda\gamma_{\mu}
\lambda_{5\alpha}\right)+\sqrt{2}
\left(\barlambda_{\delta}\gamma_{\mu}
\lambda_{\alpha}^{~~\delta}\right)\right]\no
&&+2~\overline{T}^{\mu\alpha}\left[\left(\barlambda\gamma_{\mu}
\lambda_{\alpha}\right)-\sqrt{2}
\left(\barlambda_{5\delta}\gamma_{\mu}
\lambda_{\alpha}^{~~\delta}\right)\right]\no
&&-\sqrt{2}~C^{\mu\alpha\beta}
\left[\left(\barlambda_{\alpha}\gamma_{\mu}
\lambda_{\beta}\right)+\left(\barlambda_{5\alpha}\gamma_{\mu}\lambda_{5\beta}
\right)
+2~
\left(\barlambda_{\alpha\delta}\gamma_{\mu}\lambda_{\beta}^{~~\delta}
\right)\right],
\label{LagPsi2}
\eea
In deriving Eq.(\ref{LagPsi2}), the following Majorana 
relations for anticommuting
fields $\lambda_{a}$ have been used
\be
\bar{\lambda}_a \gamma_{\mu}\lambda_b
=-\bar{\lambda}_b\gamma_{\mu} \lambda_a,~~~
\bar{\lambda}_a\gamma_{\mu}\gamma_5 \lambda_b=
\bar{\lambda}_b\gamma_{\mu}\gamma_5 \lambda_a,~~~
\bar{\lambda}_a\lambda_b=\bar{\lambda}_b\lambda_a
\label{majrules}
\ee
where in Eqs.(\ref{LagPsi2}) and (\ref{majrules}),
the definitions (\ref{majorana}) (or equivalently 
$\bar{\lambda}_a\equiv \lambda_a^t C$) are understood.
Then, due to the fact 
that the lagrangian in Eq.(\ref{LagPsi2}) does not depend 
on the adjoint fields $\lambda_a^{\dag}$ (or analogously
$\barlambda_a^{\dag}$), the hermiticity is lost \cite{nicolai}.

For completeness, we report the infinitesimal gauge transformations 
for the component fermion fields 
\bea
\delta_G \lambda&=&2\left( \lambda^{\alpha}\bar{\epsilon}_{\alpha}-
\lambda_5^{\alpha}\epsilon_{\alpha}\right),\no
\delta_G \lambda_{\alpha} &=& 2\left(-\lambda \bar{\epsilon}_{\alpha}+
\sqrt{2}~\lambda^{\beta} \epsilon_{\beta\alpha}+
\lambda_{5\alpha} \bar{\epsilon}-
\sqrt{2}~\lambda_{\beta\alpha} \epsilon^{\beta}\right),\no
\delta_G \lambda_{5\alpha} &=& 2\left(\lambda \epsilon_{\alpha}+
\sqrt{2}~\lambda_{5}^{\beta} \epsilon_{\beta\alpha}-
\lambda_{\alpha} \bar{\epsilon}-
\sqrt{2}~\lambda_{\beta\alpha} \bar{\epsilon}^{\beta}\right),\no
\delta_G \lambda_{\alpha\beta}&=&-\sqrt{2}\left(
\lambda_{\alpha}\epsilon_{\beta}+\lambda_{5\alpha}\bar{\epsilon}_{\beta}+
2\lambda_{\mu\alpha}\epsilon^{\mu}_{~\beta}\right)-
\left(\alpha \leftrightarrow \beta\right).
\label{Gtransf}
\eea
These transformations,
together with the corresponding ones for the gauge fields
(\ref{gaugetransf}), 
leave invariant the lagrangian $L_F$.

We now give the results for the Euclidean lagrangian $L^{SUSY}_{N=1}$
which is invariant (up to a total derivative)
under the N=1 (off-shell) SUSY transformations
\be
L^{SUSY}_{N=1}=L_E+L_F-
\frac{1}{2} \left(\overline{\D}^2+\D_{\mu}\D^{\mu}+
\overline{\D}_{\mu}\overline{D}^{\mu}+\D_{\mu\nu} \D^{\mu\nu}
\right)
\label{Ssusy}
\ee
Note that in (\ref{Ssusy}) the $L_E$ and $L_F$ are the ones 
given in Eq.(\ref{Lpure}) and Eq.(\ref{LagPsi1}),
respectively, and the auxiliary (bosonic) fields 
$\overline{\D}$, $\D^{\mu}$,$\overline{\D}^{\mu}$,
and $\D^{\mu\nu}$ (with $\overline{\D}$, $\D^{\mu}$,
and analogously $\overline{\D}^{\mu}$,
$\D^{\mu\nu}$, being, respectively, 
a scalar, vector, and antisymmetric tensor under $O(4)$) 
should be added to the lagrangian in order to close the off-shell 
SUSY algebra. We recall that the lagrangian 
in Eq.(\ref{Ssusy}) is not hermitean, but it satisfies 
the Osterwalder--Schrader reflection positivity \cite{nicolai}.

Finally, the (off-shell) N=1 SUSY transformations which leave invariant
the lagrangian (\ref{Ssusy}) (up to a total derivative) 
are given by
\bea
\delta_S A_{\mu}&=&\bar{\omega}\gamma_{\mu}\lambda,~~~~~~
\delta_S T_{\mu\alpha}=\bar{\omega}\gamma_{\mu}\lambda_{\alpha},\no
\delta_S \overline{T}_{\mu\alpha}&=&
\bar{\omega}\gamma_{\mu}\lambda_{5\alpha},~~
\delta_S C_{\mu\alpha\beta}=
\bar{\omega}\gamma_{\mu}\lambda_{\alpha\beta},\no
\delta_S \lambda&=&\left(\frac{\imath}{2}\sigma^{\mu\nu} 
\barF_{\mu\nu}+\gamma_5\overline{\D}\right)\omega,~~~~~~
\delta_S \lambda_{\alpha}=\left(\frac{\imath}{2}\sigma^{\mu\nu} 
F_{\mu\nu\alpha}+\gamma_5\D_{\alpha}\right)\omega,\no
\delta_S \lambda_{5\alpha}&=&\left(\frac{\imath}{2}\sigma^{\mu\nu} 
\barF_{\mu\nu\alpha}+\gamma_5\overline{\D}_{\alpha}\right)
\omega,~~
\delta_S \lambda_{\alpha\beta}=
\left(\frac{\imath}{2}\sigma^{\mu\nu} F_{\mu\nu\alpha\beta}+
\gamma_5 \D_{\alpha\beta}\right)\omega,\no
\delta_S \overline{\D} &=& 
\bar{\omega} \gamma_5 \gamma^{\mu} {\cal{D}}_{\mu} 
\lambda,~~~~~~
\delta_S \D_{\alpha} = \bar{\omega} \gamma_5 \gamma^{\mu} {\cal{D}}_{\mu} 
\lambda_{\alpha},\no
\delta_S \overline{\D}_{\alpha} 
&=& \bar{\omega} \gamma_5 \gamma^{\mu} {\cal{D}}_{\mu} 
\lambda_{5\alpha},~~
\delta_S \D_{\alpha\beta}=
\bar{\omega} \gamma_5 \gamma^{\mu} {\cal{D}}_{\mu} \lambda_{\alpha\beta},
\label{SUSYtransf}
\eea
where the complex four--spinor $\omega$ is the (gauge--singlet)
SUSY infinitesimal anticommuting parameter, 
and $\bar{\omega}$ satisfies the identity $\bar{\omega}\equiv \omega^t C$.
In the proof of supersymmetric invariance, 
the relations (\ref{majrules}) should be used.
Note that we expressed the SUSY transformations only in terms of 
the $\lambda_a$ fields, the corresponding ones for the transpose
fields $\lambda_a^t$ can be simply derived.
The expressions for the field-strength $F^a_{\mu\nu}$ are 
\bea
\barF_{\mu\nu}&=&\partial_{\mu}\bar{A}_{\nu}-
2 g \left(T_{\mu\alpha}\overline{T}_{\nu}^{~\alpha}\right)
-(\mu\leftrightarrow\nu),\no
F_{\mu\nu\alpha}&=&\partial_{\mu}T_{\nu\alpha}-
2 g \left(\sqrt{2}~T_{\mu}^{~\beta} C_{\nu\beta\alpha}
-\bar{A}_{\mu} \overline{T}_{\nu\alpha}\right)-(\mu\leftrightarrow\nu),\no
\barF_{\mu\nu\alpha}&=&\partial_{\mu}\overline{T}_{\nu\alpha}-
2 g \left(\sqrt{2}~\overline{T}_{\mu}^{~\beta} C_{\nu\beta\alpha}
+\bar{A}_{\mu} T_{\nu\alpha}\right)-(\mu\leftrightarrow\nu),\no
F_{\mu\nu\alpha\beta}&=&\partial_{\mu}C_{\nu\alpha\beta}+
\sqrt{2} g \left(T_{\mu\alpha} T_{\nu\beta}+\barT_{\mu\alpha} \barT_{\nu\beta}
-2C_{\mu\beta\delta}
C_{\nu\alpha}^{~~~\delta}\right)-(\mu\leftrightarrow\nu).
\label{fieldtrengths}
\eea
The covariant derivatives are defined as
${\cal D}_{\mu} \lambda^a=\partial_{\mu}\lambda^a+g \Delta_{\mu} \lambda^a$,
where the expressions for 
$\Delta_{\mu} \lambda^a$ are
\bea
\Delta_{\mu} \lambda&=&2\left( \lambda^{\alpha}\overline{T}_{\mu\alpha}-
\lambda_5^{\alpha}T_{\mu\alpha}\right),\no
\Delta_{\mu} \lambda_{\alpha} &=& 2\left(-\lambda \overline{T}_{\mu\alpha}+
\sqrt{2}~\lambda^{\beta} C_{\mu\beta\alpha}+
\lambda_{5\alpha} \bar{A}_{\mu}-
\sqrt{2}~\lambda_{\beta\alpha} T_{\mu}^{~\beta}\right),\no
\Delta_{\mu}\lambda_{5\alpha} &=& 2\left(\lambda \overline{T}_{\mu\alpha}+
\sqrt{2}~\lambda_5^{\beta} C_{\mu\beta\alpha}-
\lambda_{\alpha} \bar{A}_{\mu}-
\sqrt{2}~\lambda_{\beta\alpha} \overline{T}_{\mu}^{~\beta}\right),\no
\Delta_{\mu} \lambda_{\alpha\beta}&=&-\sqrt{2}\left(
\lambda_{\alpha}T_{\mu\beta}+\lambda_{5\alpha}\overline{T}_{\mu\beta}+
2\lambda^{\delta}_{~\alpha}C_{\mu\delta\beta}\right)-
\left(\alpha \leftrightarrow \beta\right).
\eea

The lagrangian (\ref{LagPsi2}) is not written in terms of the 
$O(4)$ irreducible components, 
but this can be easily obtained by taking into account 
the decompositions (\ref{IRdecomp}), (\ref{Psidecomp}), and the relations
(\ref{majrules}).
We observe that non--trivial couplings between the irreducible
representation of the gauge and fermion fields can arise in the interacting 
lagrangian $L_F^I$.
After some straightforward algebraic manipulations,
in terms of the fermion irreducible representations (\ref{Psidecomp}),
the free Lagrangian $L_F^0$ is, up to total derivatives,
\bea
L_F^0&=&\frac{1}{2}\bar{\lambda} \gamma^{\mu}\partial_{\mu}\lambda
+\frac{1}{2}\bar{\psi}_{\alpha}\gamma^{\mu}\partial_{\mu}\psi^{\alpha}
+\frac{1}{2}\bar{\psi}\gamma^{\mu}\partial_{\mu}\psi
-\sqrt{2}\bar{\psi}\partial^{\mu}\psi_{\mu}
+\frac{1}{2}
\bar{\psi}_{5\alpha}\gamma^{\mu}\partial_{\mu}\psi_5^{\alpha}\no
&+&\frac{1}{2}\bar{\psi}_5\gamma^{\mu}\partial_{\mu}\psi_5
-\sqrt{2}\bar{\psi}_5\gamma_5\partial^{\mu}\psi_{5\mu}
+\frac{1}{2}\bar{\psi}_{\mu\nu}\gamma^{\alpha}\partial_{\alpha}\psi^{\mu\nu}
-2\bar{\xi}^{\nu}\left(\partial_{\nu}\xi
+\partial^{\mu}\psi_{\mu\nu}\right),
\label{LF0}
\eea
where, in deriving the expression (\ref{LF0}),
the relations (\ref{majrules}) were used, and the following definitions hold
\be
\lambda \equiv C\bar{\lambda}^{t},
~~\psi_{\mu}\equiv C\bar{\psi}^{t}_{\mu},~~
\psi_{5 \mu}\equiv C\bar{\psi}^{t}_{5 \mu},~~
\psi_{\mu\nu}\equiv C\bar{\psi}^{t}_{\mu\nu},~~
\xi_{\mu}\equiv C\bar{\xi}^{t}_{\mu}~.
\label{majoranacomp}
\ee
The corresponding gauge or SUSY transformations for the $O(4)$ 
irreducible representations (see the r.h.s. of Eq.(\ref{Psidecomp})
\footnote{
We recall  that in order to make a comparison 
(by using the notation of (\ref{IRRtransf}))
between Eqs. (\ref{Psidecomp}) and 
(\ref{IRRtransf}), the
``bar'' over the $\bar{\gamma}_{\mu}$ matrices in 
(\ref{Psidecomp}) should be eliminated.})
can now be simply obtained by projecting the 
reducible transformations
$\delta \lambda^{\mu}$, $\delta \lambda^{\mu\nu},~\dots$ etc., defined in
Eqs.(\ref{Gtransf}) or (\ref{SUSYtransf}), into the following ones
\bea
\delta \psi&=&\frac{1}
{2\sqrt{2}}\gamma_{\mu} \left(\delta \lambda^{\mu}\right),~~~~~~~~
\delta \psi_5=\frac{1}{2\sqrt{2}}
\gamma_{\mu}\left(\delta \lambda^{5\mu}\right),\no
\delta\psi^{\mu}&=&\delta \lambda^{\mu}-\frac{1}
{\sqrt{2}}\gamma_{\mu}\left(\delta \psi\right),~~
\delta\psi_5^{\mu}=\delta 
\lambda_5^{\mu}-\frac{1}{\sqrt{2}}\gamma_{\mu}
\left(\delta \psi_5\right),\no
\delta \xi&=&-\frac{\imath}{6}
\sigma_{\mu\nu} \left(\delta \lambda^{\mu\nu}\right),~~~~~~~~~~~
\delta \xi_{\mu}=\gamma_{\mu}
\left[\delta \lambda^{\mu\nu}-
\frac{\imath}{2}\sigma^{\mu\nu} \left(\delta \xi\right)\right],\no
\delta \psi^{\mu\nu}&=&
\delta \lambda^{\mu\nu}-\frac{\imath}{2} 
\sigma^{\mu\nu} \left(\delta \xi\right) -\frac{1}{4}
\left[\gamma^{\mu} 
\left(\delta \xi^{\nu}\right)-\gamma^{\nu} \left(\delta \xi^{\mu}\right)
\right].
\label{IRRtransf}
\eea

We now analyze the N=2 SUSY extension of the model in
\cite{eg} by requiring, as in the Zumino model \cite{zumino},
the hermiticity of the action in Euclidean space.
In order to obtain an hermitean action, one has to introduce a complex 
fermionic field $\hat{\Psi}$ and its adjoint one, namely $\bar{\hat{\Psi}}$.
The  $\hat{\Psi}$ and $\bar{\hat{\Psi}}$
have, respectively, the $O(4)$ coordinate transformation properties as 
defined in (\ref{Psixtransf}) and (\ref{Psibarxtransf}).
For our convenience we formally decompose $\hat{\Psi}$ and $\bar{\hat{\Psi}}$
as in Eq.(\ref{psibardecomp}) 
and Eq.(\ref{psidecomp}), respectively.
The main difference with the previous N=1 extension is that, in 
order to obtain an hermitean action, the 
component fields of $\bar{\hat{\Psi}}$ (namely 
$\bar{\lambda}_a$ in (\ref{psibardecomp})), are the corresponding
adjoint ones of the $\lambda_a$ fields. Obviously
the definitions (\ref{majorana}) and the relations 
(\ref{majrules}), unlike the N=1 case, do not apply in this case.

The fermionic degrees of freedom are doubled with
respect to the $N=1$ theory in Eq.(\ref{Ssusy}), and so
new bosonic (real) matter fields $\hat{H}$ and $\hat{W}$
should be introduced in order to close the N=2 SUSY algebra \cite{zumino}.
In components we have
\bea
\hat{H}&=&
\bar{H}\gamma_5+H_{\mu}\gamma^{\mu}+\bar{H}_{\mu}\gamma^{\mu}\gamma_5+
H_{\mu\nu}\frac{\sigma^{\mu\nu}}{\sqrt{2}}\no
\hat{W}&=&
\bar{W}\gamma_5+W_{\mu}\gamma^{\mu}+\bar{W}_{\mu}\gamma^{\mu}\gamma_5+
W_{\mu\nu}\frac{\sigma^{\mu\nu}}{\sqrt{2}}
\label{N2scalars}
\eea
where $\hat{H}^{\dag}=\hat{H}$ and $\hat{W}^{\dag}=\hat{W}$.
Under the $O(4)$ coordinate transformations and extended gauge transformations,
the field $\hat{H}$ (and analogously for $\hat{W}$) should 
transform as follows
\be
\hat{H}(x)\to \hat{H}^{\prime}(x')=
S(\Lambda)\hat{H}(x)S^{-1}(\Lambda)~,~~~~
\hat{H}(x)\to \hat{H}^{G}(x)=U(x)\hat{H}(x)U^{-1}(x)
\label{Htransf}
\ee
Unlike the Zumino model where the gauge symmetry is internal,
the components
$\bar{H}$, $H_{\mu}$ (and analogously for $\bar{H}_{\mu}$),
and $H_{\mu\nu}$ transform, according to Eq.(\ref{Htransf}),
as a scalar, vector, and antisymmetric tensor of rank 2
under $O(4)$ coordinate rotations.
Analogous considerations hold for the $\hat{W}$ field.

Finally, the expression (in compact notation) for the 
N=2 SYM lagrangian, which is invariant (up to a total derivative) 
under on-shell SUSY transformations and $SU(4)$
extended gauge transformations, is given by \cite{zumino}
\bea
L^{SUSY}_{N=2}&=&L_E+
\frac{\imath}{8}Tr\left[\bar{\hat{\Psi}}\bar{\gamma}_{\mu}(
\hat{D}^{\mu}\hat{\Psi})-
(\hat{D}^{\mu}\bar{\hat{\Psi}})\bar{\gamma}_{\mu}\hat{\Psi}\right] \no
&+& \frac{1}{8}Tr\left[\left(\hat{D}_{\mu} \hat{H}^{\mu}\right)^2\right]-
\frac{1}{8}Tr\left[\left(\hat{D}_{\mu} \hat{W}^{\mu}\right)^2\right]\no
&-&\frac{\imath g}{4}Tr\left\{\bar{\hat{\Psi}}
\left[\left(\hat{W}-\bar{\gamma_5} \hat{H}\right)~,~
\hat{\Psi}\right]\right\}
+\frac{g^2}{8}Tr\left\{\left[\hat{W}~,~\hat{H}\right]^2\right\}
\label{LsusyN2}
\eea
where the covariant derivative $D_{\mu}$ is defined in Eq.(\ref{covderiv}),
the commutator $[~,~]$ and the trace $Tr$ are taken in the Clifford
algebra basis $\Gamma_i$. 
We recall that 
the $\bar{\gamma}_{\mu}$ and $\bar{\gamma}_5$ matrices appearing in
Eq.(\ref{LsusyN2}), belong to the other Clifford basis $\bar{\Gamma}_i$
defined in Eq.(\ref{Gammabasis}).
It is easy to prove that the lagrangian (\ref{LsusyN2}) is hermitean and
invariant under the following on--shell N=2 SUSY transformations
\cite{zumino}
\bea
\delta \hat{W}&=&\imath \bar{\alpha} \hat{\Psi}-\imath
\bar{\hat{\Psi}}\alpha~,~~~~
\delta \hat{H}=\imath \bar{\alpha} \bar{\gamma}_5 
\hat{\Psi}-\imath \bar{\hat{\Psi}} \bar{\gamma_5} \alpha~,~~~~
\delta \hat{A}_{\mu}=\imath \bar{\alpha}\bar{\gamma}_{\mu}\hat{\Psi}-
\imath \bar{\hat{\Psi}}\bar{\gamma}_{\mu}\alpha\no
\delta \hat{\Psi}&=&\frac{\imath}{2}\hat{F}_{\mu\nu}\bar{\sigma}^{\mu\nu} 
\alpha+\left(D_{\mu} \hat{W}
+\bar{\gamma}_5D_{\mu}\hat{H}\right)\bar{\gamma}_{\mu}\alpha
-\imath \left[\hat{W}~,~\hat{H}\right]\bar{\gamma}_5\alpha\no
\delta \bar{\hat{\Psi}}&=&\frac{\imath}{2}
\bar{\alpha}\hat{F}_{\mu\nu}\bar{\sigma}^{\mu\nu} 
+\bar{\alpha}\bar{\gamma}_{\mu}\left(D_{\mu} \hat{W}
+\bar{\gamma}_5D_{\mu}\hat{H}\right)
-\imath \bar{\alpha}\bar{\gamma}_5\left[\hat{W}~,~\hat{H}\right]
\label{N2susytransf}
\eea
where complex four-spinor $\alpha$ is the (gauge--singlet) 
infinitesimal parameter of the
N=2 SUSY transformations, and $\bar{\alpha}$ is its adjoint one.
Note that, unlike the N=1 case, the presence of both SUSY parameters
$\alpha$ and $\bar{\alpha}$, ensures that the N=2 SUSY 
transformations are hermitean. Clearly,
the number of fermionic degrees of freedoms has doubled with
respect to the N=1 case.\\
We do not give the corresponding results of 
Eqs.(\ref{LsusyN2}) and (\ref{N2susytransf})
in terms of the $O(4)$ irreducible components, since their expressions
are quite complicate. However these can be easily obtained 
by applying the standard decomposition methods (into 
$O(4)$ irreducible representations) 
explained in section [2] and in the present one.

We stress that the $N=2$ SYM theory in Eq.(\ref{LsusyN2}) 
is particularly interesting,
since it naturally contains matter scalar fields ($\bar{H}$
and $\bar{W}$)
which could generate the spontaneous symmetry breaking of the extended gauge 
symmetry.
We will return on this point in section [7], where 
the spontaneous symmetry breaking will be analyzed. 
Moreover, it would be interesting to study, in Minkowski space,
the non--perturbative solutions of the N=2 SYM model in 
Eq.(\ref{LsusyN2}), by generalizing the corresponding ones of
Seiberg--Witten theory \cite{SW}.

Before concluding the present section we underline
the characterizing properties of the above introduced 
supersymmetric extensions.
The extended gauge symmetry mixes fields of different
spin but with the same statistics. On the contrary a SUSY transformation mixes
fields of different statistic.
Moreover, due to the fact that the maximum spin of the gauge-charges is spin-1 
and the SUSY charges have spin-1/2, the transformations which leave invariant
the total action, can mix fields of spin-S 
with fields of spin-($S\pm\Delta S$), where $\Delta S = 1$ and 
$\Delta S= 1/2$ correspond to a gauge and SUSY transformation, respectively.
Finally the product of a SUSY and gauge transformation can 
produce a $|\Delta S| = 3/2$ spin-transition.
\section{Internal Symmetries}
In this section we generalize the model in \cite{eg}
so as to include the standard internal gauge symmetries.
This is effected by restricting our choice to the compact groups SU(N).

This generalization is obtained by considering a Lie group whose algebra
is given by the tensorial product of the Clifford and $U(N)$ algebra bases.
It is important to note that the elements given by 
the tensorial product of the unity matrix of $SU(N)$ with the
Clifford algebra basis are necessary in order to close the algebra. 
As a consequence we have (in 4 dimensions) a Lie group containing
$16\times N^2 -1$ generators,
which form the algebra (in Euclidean space) of $SU(4\times N)$.
The corresponding elements of the extended algebra (which we call $Y$)
containing internal symmetries generators 
can be represented in the compact form
\be
Y_i^{{\bf a}}=\left\{\tilde{\Gamma}_i\ptensor {\unityT},
~\tilde{\Gamma}_i\ptensor {\bf T^a},
{\unityG}\ptensor {\bf T^a} \right\},
\label{Ygener}
\ee
where the symbol $\ptensor$ indicates the
standard tensorial product between the matrices $\Gamma_i$ 
(of the Clifford algebra basis) and ${\bf T^a}$ (of $SU(N))$,
$\unityG$, and $\unityT$ are the unity matrix of $\Gamma_i$ and
$SU(N)$ respectively. 
By $\tilde{\Gamma}_i$ we denote any element of 
the reduced Clifford algebra basis which does not contain $\unityG$.
The matrices ${\bf T^a}$ satisfy 
the following commutation and anticommutation rules
\be
[{\bf T^a},{\bf T^b}]=
i f^{{\bf abc}} {\bf T^c},~~~\{{\bf T^a,T^b}\}=
2 C_F\delta^{{\bf ab}}{\unityT}+d^{{\bf abc}} {\bf T^c},~~
[{\bf T^a},\Gamma_i]=0,
\label{Talgebra}
\ee
where the following normalizations are used
$Tr({\bf T^a T^b})=1/2\delta^{{\bf a b}}$ 
and $C_F=1/(2N)$ for the ${\bf T^a}$ matrices in
the fundamental representation. The structure function $d^{{\bf abc}}$ is 
a complete symmetric tensor with null traces $d^{{\bf abc}} \delta_{{\bf ab}}=0$. 

This $Y$ algebra was proposed a long time ago 
in Refs.\cite{su12a},\cite{su12b}  in the context
of the relativistic extensions of the $SU(6)$ model \cite{su6}
for the strong interactions
(the corresponding Lie structure functions can be found in Ref.\cite{su12a}).
We want to stress that, although there is 
a coincidence of the symmetry group, 
there is no analogy between the model in \cite{eg} and the model proposed 
in Refs.\cite{su12a},\cite{su12b}, and the theoretical inconsistencies 
\cite{nogoth} which are present in the latter do not affect 
the theory of the model in \cite{eg}.

Now we give the generalization of the  gauge transformations 
(\ref{Agtransf})-(\ref{eps}) by including the internal symmetries
generated by the group $SU(4\times N)$.
We begin with the 
unitary gauge group element ${\bf U}(x)$ which is given by
\bea
{\bf U}(x)&=&\exp \imath\left\{
\hat{\epsilon}\ptensor \unityT+
\epsilon^{{\bf a}}(x)\left({\bf T^a}\ptensor\unityG\right)+
\bar{\epsilon}^{{\bf a}}(x)\left({\bf T^a}\ptensor\gamma_{5}\right)+
\epsilon_{\alpha}^{{\bf a}}(x)
\left({\bf T^a}\ptensor\gamma^{\alpha}\right)\right.\no
&+&\left.
\bar{\epsilon}_{\alpha}^{{\bf a}}(x)\left({\bf T^a}\ptensor
i\gamma^{\alpha}\gamma_{5}\right)
+\epsilon^{{\bf a}}_{\alpha\beta}(x)\left({\bf T^a}\ptensor
\frac{\sigma^{\alpha\beta}}{\sqrt{2}}\right)
\right\},
\label{Uint}
\eea
where $\hat{\epsilon}$ is given in Eq.(\ref{eps}).
We require that 
${\bf U}(x)$ transforms as $U(x)$ in Eq.(\ref{Uxtransf}) under 
coordinate transformations. Then, due to Eq.(\ref{Srepr}) 
and to the fact that the $SU(N)$ generators commute with the $O(4)$
transformations $[{\bf T^a},S(\Lambda)]=0$, only the ``greek'' indices
in the $\epsilon_i$ parameters in (\ref{Uint}), 
transform like vectorial indices under $O(4)$ coordinate transformations.
\\
For the generalization of the gauge potential ${\bf \hat{A}}_{\mu}$ 
we may proceed in the same way. 
We define ${\bf \hat{A}}_{\mu}$  to have the same
coordinate and gauge transformation properties of
$\hat{A}_{\mu}$ in (\ref{Axtransf}) and (\ref{Agtransf}) respectively.
Then we decompose ${\bf \hat{A}}_{\mu}$
along the $Y$ algebra basis in  (\ref{Ygener})
as follows
\footnote{
To avoid confusion note that 
the symbols ${\bf T^a}$ and $T^{{\bf a}}_{\mu\nu}$ indicate
the $SU(N)$ generators and the tensor fields along the $\gamma_{\nu}
{\bf T^a}$ basis respectively.}
\bea
{\bf \hat{A}}_{\mu}&=&\hat{A}_{\mu}\otimes{\unityT}+
A^{{\bf a}}_{\mu}(x)\left({\bf T^a}\otimes {\unityG}\right)+
\bar{A}^{{\bf a}}_{\mu}(x)\left({\bf T^a}\gamma_{5}\otimes {\unityG}\right)+
T_{\mu\alpha}^{{\bf a}}(x)\left({\bf T^a}\otimes\gamma^{\alpha}\right)\no
&+&\bar{T}_{\mu\alpha}^{{\bf a}}(x)\left({\bf T^a}\otimes 
i\gamma^{\alpha}\gamma_{5}\right)
+C^{{\bf a}}_{\mu\alpha\beta}(x)
\left({\bf T^a}\otimes\frac{\sigma^{\alpha\beta}}{\sqrt{2}}\right),
\label{AIS}
\eea
where the expression of $\hat{A}_{\mu}$ is given in Eq.(\ref{Adecomp}).
As in the case of $\hat{A}_{\mu}$, the ``greek'' indices 
of the component fields in (\ref{AIS})
transform like vectorial indices under the $O(4)$ coordinate transformations.

The pure YM lagrangian in terms of the
$\bf{\hat{A}}_{\mu}$ field which 
generalizes (\ref{Lpure}), so as to include the 
local internal symmetries, has the same formal expression of (\ref{Lpure}) 
\bea
{\bf L^{IS}}(x)&=&{1\over{8}}Tr[{\bf\hat{F}}_{\mu\nu}{\bf \hat{F}}^{\mu\nu}],
\no
{\bf \hat{F}}_{\mu\nu}(x)&=&\partial_{\mu}{\bf \hat{A}}_{\nu}
-\partial_{\nu}{\bf \hat{A}}_{\mu}+\imath g[{\bf \hat{A}}_{\mu},
{\bf \hat{A}}_{\nu}],
\label{LagIS}
\eea
where the commutators and trace have to be evaluated on the $Y$ algebra.
The full expression for the lagrangian ${\bf L^{IS}}$
in terms of the component fields in (\ref{AIS})
as well as the 
infinitesimal gauge transformations which leave invariant ${\bf L^{IS}}$
are given in the Appendix.

It is worth noting that, since the $Y$ algebra contains also
the anticommutators of the $SU(N)$ generators,  the couplings between 
the extended $\hat{A}_{\mu}$ gauge fields (\ref{Adecomp})
and the corresponding ones $\hat{A}_{\mu}^{\bf a}$ 
(along the ${\bf T^a}$ basis) appear in the
lagrangian (\ref{LagIS1}-\ref{LagIS2}) (see Appendix).
As a consequence, in the gauge lagrangian the complete 
symmetric structure functions $d^{\bf abc}$ of the $SU(N)$ gauge group
appears. The presence of the  $d^{\bf abc}$ structure functions 
is one of the most interesting aspect of this generalized model,
which has no precedent (as far as we know) in any known gauge theory.
Moreover it is remarkable to note that, after rescaling in 
Eqs.(\ref{LagIS1}-\ref{LagIS2})  the kinetic term 
(by setting it in the canonical form)
of the pure extended gauge fields $\hat{A}_{\mu}$,
the couplings between the $\hat{A}_{\mu}$ and  
$\hat{A}_{\mu}^{\bf a}$ gauge fields 
decrease as $g/\sqrt{N}$ (in the fundamental representation) 
in the large $N$ limit.

The coupling of the gauge field ${\bf \hat{A}}_{\mu}$ 
with fermion fields  in the adjoint 
representation can be obtained in a way similar to the one 
used in section [4] 
by taking into account the different commutation rules of the $Y$ algebra.
Typically the generalization so as to include
the internal symmetries in the SUSY transformations (\ref{IRRtransf})
is also straightforward.

\section{Bosonic Matter Fields}
In this section we study the couplings of the
gauge fields with the matter fields $\Sigma$ 
which transform as the fundamental 
representation of the extended $U(4)$ gauge group.
We will see that for these kind of fields 
the only consistent gauge--invariant theories, which 
are compatible with the spin--statistic theorem \cite{weinb},  
are of bosonic type.
In particular, in order to build a coordinate-- and 
$U(4)$ gauge--invariant lagrangian, the 
field $\Sigma$ of lowest spin must be a non--hermitean one $\Sigma_{ij}$
(where $i$ and $j$ are spinorial indices) which transforms, under $O(4)$
coordinate transformations, as the adjoint of the spinorial 
representation of the rotation group, namely
\bea
x\to x^{\prime}_{\mu}&=&\Lambda_{\mu}^{~\nu} x_{\nu},\no
\Sigma_{ij}(x)\to \Sigma_{ij}^{\prime}(x^{\prime})&=&
(S(\Lambda)\Sigma(x)S^{-1}(\Lambda))_{ij},
\label{Sigmaxtransf}
\eea
where $S(\Lambda)$ is defined in Eq.(\ref{Srepr}) 
and the standard matrix multiplication is assumed.
Therefore the $\Sigma$ field can be decomposed as follows
\be
\Sigma=\unityG \varphi +\gamma_5 \bar{\varphi} +
\gamma_{\mu} B^{\mu}+\gamma_{\mu}\gamma_5 \bar{B}^{\mu}+
\sigma_{\mu\nu} C^{\mu\nu},
\label{Sigmadecomp}
\ee
where $\varphi (\bar{\varphi})$, 
$B_{\mu} (\bar{B}_{\mu})$ and $C_{\mu\nu}$ are complex scalar, vectorial 
and tensorial fields respectively.
Now we define its properties under local gauge transformations. In particular
we have
\be
\Sigma\to \Sigma^{G}= U(x)\Sigma,~~~
\Sigma^{\dag}\to \Sigma^{\dag G}=\Sigma^{\dag}U^{\dag}(x),
\label{SigmaGtransf}
\ee
where $U(x)$ is given in Eq.(\ref{Udecomp}) and the standard multiplication
between matrices is assumed.
Note that the group $U(4)$ is defined to act on $\Sigma$ and $\Sigma^{\dag}$ 
on the {\it right} and {\it left} respectively.
Finally, the most general gauge--invariant and renormalizable 
(by power counting) lagrangian, which contains only the $\Sigma$ field and the
gauge connection $\hat{A}_{\mu}$, is given by
\bea
L_{\Sigma}&=&
Tr\left[\Sigma^{\dag} D_{\mu} D^{\mu}
\Sigma\right]+
Tr\left[\gamma_5 \Sigma^{\dag} D_{\mu} D^{\mu}
\Sigma\right]+Tr\left[\sigma^{\mu\nu}
\Sigma^{\dag}\left[ D_{\mu}, D_{\nu}
\right]\Sigma\right]+\no
&&Tr\left[\tilde{\sigma}^{\mu\nu}
\Sigma^{\dag}\left[ D_{\mu}, D_{\nu}
\right]\Sigma\right]+\imath
m_1 Tr\left[\gamma^{\mu}\Sigma^{\dag} D_{\mu}
\Sigma\right]+
m_2 Tr\left[\gamma^{\mu}\gamma_5\Sigma^{\dag} D_{\mu}
\Sigma\right]+\no
&&m_3^2 Tr\left[
\Sigma^{\dag}\Sigma\right] + m_4^2 Tr\left[
\gamma_5\Sigma^{\dag}\Sigma\right]+
\lambda_1 Tr\left[(\Sigma^{\dag}\Sigma)^2\right] +
\lambda_2 Tr\left[\gamma_5(\Sigma^{\dag}\Sigma)^2\right] +\no
&&\lambda_3 Tr\left[\gamma_5\Sigma^{\dag}\Sigma\gamma_5
\Sigma^{\dag}\Sigma\right] +
\lambda_4 Tr\left[\gamma_{\mu}\Sigma^{\dag}\Sigma
\gamma^{\mu}\Sigma^{\dag}\Sigma\right] +
\lambda_5 Tr\left[\gamma_{\mu}\gamma_5\Sigma^{\dag}\Sigma
\gamma^{\mu}\gamma_5\Sigma^{\dag}\Sigma\right] +\no
&&\lambda_6 Tr\left[\gamma_{\mu}\Sigma^{\dag}\Sigma
\gamma^{\mu}\gamma_5\Sigma^{\dag}\Sigma\right] +
\lambda_7 Tr\left[\sigma_{\mu\nu}\Sigma^{\dag}\Sigma
\sigma^{\mu\nu}\Sigma^{\dag}\Sigma\right] +
\lambda_8 Tr\left[\tilde{\sigma}_{\mu\nu}\Sigma^{\dag}\Sigma
\sigma^{\mu\nu}\Sigma^{\dag}\Sigma\right]
\no
&&\lambda_9 Tr\left[\Sigma^{\dag}\Sigma\right]^2 +
\lambda_{10} Tr\left[\gamma_5\Sigma^{\dag}\Sigma\right]^2+
\lambda_{11} Tr\left[\Sigma^{\dag}\Sigma\right]
Tr\left[\gamma_5\Sigma^{\dag}\Sigma\right] +\no
&&\lambda_{12} Tr\left[\gamma_{\mu}\Sigma^{\dag}\Sigma\right]
Tr\left[\gamma^{\mu}\Sigma^{\dag}\Sigma\right] +
\lambda_{13} Tr\left[\gamma_{\mu}\gamma_5\Sigma^{\dag}\Sigma\right]
Tr\left[\gamma^{\mu}\gamma_5\Sigma^{\dag}\Sigma\right] +\no
&&\lambda_{14} Tr\left[\gamma_{\mu}\Sigma^{\dag}\Sigma\right]
Tr\left[\gamma^{\mu}\gamma_5\Sigma^{\dag}\Sigma\right] +
\lambda_{15} Tr\left[\sigma_{\mu\nu}\Sigma^{\dag}\Sigma\right]
Tr\left[\sigma^{\mu\nu}\Sigma^{\dag}\Sigma\right] + \no
&&\lambda_{16} Tr\left[\tilde{\sigma}_{\mu\nu}\Sigma^{\dag}\Sigma\right]
Tr\left[\sigma^{\mu\nu}\Sigma^{\dag}\Sigma\right] +
h.c.,
\label{Lsigma}
\eea
where the trace $Tr$ is defined on the spinorial indices, 
$\tilde{\sigma}_{\mu\nu}\equiv \epsilon_{\mu\nu\alpha\beta}\sigma^{\mu\nu}$,
and the covariant derivative $D_{\mu}$
\be
 D_{\mu}^{ij}\equiv  \partial_{\mu}
\delta^{ij}+\imath g\hat{A}^{ij}_{\mu}
\label{Dcov}
\ee
transforms, according to Eq.(\ref{Agtransf}), as $U(x)D_{\mu}U^{\dag}(x)$.
Each single term in Eq.(\ref{Lsigma}) is invariant under $O(4)$ coordinate 
rotations and $U(4)$ gauge transformations (\ref{SigmaGtransf}).
The independent coefficients $m_{i}$ ($i$=1--4) and $\lambda_{j}$ 
($j$=1--16) are mass parameters and adimensional coupling 
constants respectively.\\
It is worth to note that in Eq.(\ref{Lsigma}), apart from the first term and 
the other ones proportional to $m_3$, $\lambda_1$, and $\lambda_9$ couplings, 
the other gauge--invariant terms
have no counterpart (as far as we know) in any known gauge
theory.\footnote{
Note that in Eq.(\ref{Lsigma}) the first term and 
the other ones proportional to $m_3$, $\lambda_1$, and $\lambda_9$
couplings 
are also invariant, in addition to the (\ref{SigmaGtransf}),
under the local gauge transformations $\Sigma\to U\Sigma U^{\dag}$.}
Moreover we find that a minimal non-trivial 
lagrangian can be obtained by taking in Eq.(\ref{Lsigma}), for example,
just only the terms proportional to $m_1$ and $m_3$. 
In particular it is not difficult to see that the lagrangian 
\be
L^{min}_{\Sigma}=
\imath 
Tr\left[\gamma^{\mu}\Sigma^{\dag} D_{\mu}\Sigma\right]
+ {\rm m} Tr\left[\Sigma^{\dag}\Sigma\right] +h.c.
\label{Lfreesigma}
\ee
generates non-trivial dynamics for the component fields in $\Sigma$.
In particular one can see that, on-shell, the $B_{\mu}$ and $\bar{B}_{\mu}$
fields (defined in Eq.(\ref{Sigmadecomp})) are the dynamical ones, 
while $C_{\mu\nu}$, $\varphi$, and $\bar{\varphi}$ play the role of 
auxiliary fields and can be eliminated by solving the corresponding 
algebraic equations.

It should by now be clear that it is not possible to couple
directly the extended gauge symmetry on the ordinary spin-1/2 
fermion fields. 
Simply because the free lagrangian of a standard spin-1/2 field is not
invariant under global extended gauge transformations $U(4)$.
However it is possible to couple directly the 
matter fields $\Sigma$ to the ordinary gauge fields.
This is now shown with an example.
We assume that the $\Sigma$ are charged under
an abelian  $U(1)$ interaction.
Then we have to add in the covariant derivative $D_{\mu}$ of 
Eq.(\ref{Dcov}) the $U(1)$ gauge connection $A_{\mu}$ as follows
\be
D_{\mu }^{ij}=
\delta^{ij}\partial_{\mu}+\imath g \hat{A}_{\mu}^{ij}+
\imath e A_{\mu}\delta^{ij},
\ee
where $A_{\mu}$ is the $U(1)$ gauge field and
$e$ is the corresponding gauge coupling.
We also assume that the standard matter fields are charged under 
the $U(1)$ gauge symmetry.
In order to have a reasonable decoupling between the extended gauge sector
and the ordinary matter fields, the $\Sigma$ should have 
a mass scale much higher than ordinary matter fields scales.
In particular the decoupling between ordinary particles and 
the massless gauge or gaugino sector of the extended gauge theory is generated
when the massive $\Sigma$ are integrated out.
Therefore, effective couplings between the sectors of the
ordinary gauge and the extended gauge fields 
can be generated by means of higher dimensional
operators which are suppressed by the corresponding inverse powers of the 
typical mass scale of the $\Sigma$ fields.
This mechanism to generate couplings
between the observable sector and the light one of the 
extended gauge symmetry has some analogies with the gauge-mediated
mechanisms proposed in soft SUSY breaking models \cite{GM}.
We stress that this is one of the possible mechanisms to couple
the ordinary matter and gauge
sector to the extended one. In particular the role of the
electromagnetic gauge vector mediation described above
could be played by any other field which 
is a singlet under the extended gauge transformations.

\section{Spontaneous symmetry breaking}
In this section we analyze the 
spontaneous symmetry breaking (SSB) of the extended gauge symmetry
by means of the standard Higgs mechanism.\footnote{
It is clear that the physical interpretation of the 
Higgs mechanism makes sense only in Minkowski space, although
this mechanism can also be analyzed in Euclidean space, but, of course,
with a different meaning.}
As usual, one has to introduce bosonic matter fields with 
an {\it ad hoc} (gauge invariant) potential in order to spontaneously 
break this symmetry.
It is worth observing that the SSB of the extended
gauge symmetry 
can generate the spontaneous breaking of the $O(4)$ (coordinate) 
rotational one.
Indeed non--zero vacuum expectation values can be generated for 
some vector or tensor components of the matter fields.
Nevertheless, as we will show in the following example, there exists a special 
SSB pattern which leaves unbroken the $O(4)$ rotational symmetry.

At this purpose we introduce one
real matter field $\Phi$ in the adjoint representation
of the extended $SU(4)$ gauge group, like the fields 
$\hat{H}$ and $\hat{W}$ which appear in Eqs.(\ref{N2scalars}) 
in the N=2 SUSY model. 
Its expression in components is given by
\be
\Phi=\bar{\phi}\gamma_5+
\phi_{\mu}\gamma^{\mu}+\bar{\phi}_{\mu}\gamma^{\mu}\gamma_5+
\phi_{\mu\nu}\frac{\sigma^{\mu\nu}}{\sqrt{2}}
\label{phi}
\ee
where $\Phi^{\dag}=\Phi$, and it has
the same transformations rules of $\hat{H}$ in Eqs.(\ref{Htransf}).
We now add to its kinetic term 
the $SU(4)$ gauge invariant potential $V(x)$. The expression for its
gauge invariant lagrangian $L_{\Phi}$ is given by
\be
L_{\Phi}=\frac{1}{4}
Tr\left[\left(\hat{D}_{\mu}\Phi\right)^2\right]-V(x)~,~~~~
V(x)=\lambda
\left\{\frac{1}{4}Tr\left(\Phi^2\right)-\frac{\mu^2}{\lambda}\right\}^2
\label{VEV}
\ee
where $\lambda > 0$ 
and $\mu^2 >0$ are the self-interacting coupling constant and 
a mass term respectively. The solution for the
minimum of $V(x)$ is given by
$\langle Tr\left(\Phi^2\right)\rangle =\mu^2/\lambda$, or in component fields
\be
\langle 
\bar{\phi}^2+\phi_{\alpha}\phi^{\alpha}+\bar{\phi}_{\alpha}\bar{\phi}^{\alpha}+
\phi_{\alpha\beta}\phi^{\alpha\beta} \rangle =\frac{\mu^2}{\lambda}
\label{minimum}
\ee
Unlike the case of matter fields
in the fundamental representation, there exist 
several solutions of Eq.(\ref{minimum}) 
which correspond to non--equivalent SSB patterns \cite{Li}.
However, 
we are interested in the special ones which leave the $O(4)$
rotational group unbroken. Clearly, due to the presence of the $O(4)$
scalar $\bar{\phi}$ in the multiplet (\ref{phi}),
the only solution which satisfies this requirement is given by:
$\langle \bar{\phi}\rangle =\sqrt{\mu^2/\lambda}$, with
$\langle \bar{\phi}_{\alpha}\rangle=\langle \bar{\phi}_{\alpha}\rangle=
\langle \bar{\phi}_{\alpha\beta}\rangle=0$.
In particular the $SU(4)$ gauge group is broken, 
while the $O(4)$ rotational symmetry remains exact.\footnote{
It is worth noting that in smaller extended gauge theories, in particular
the ones with $SO(5)$ and $SO(4)$ gauge groups
(whose generators are given  by $\left\{\gamma_{\mu},~\sigma_{\mu\nu}\right\}$ 
and $\sigma_{\mu\nu}$ respectively), there are no
SSB patterns in the adjoint representation which would preserve the 
$O(4)$ rotational symmetry. This is 
connected to the fact that for these smaller groups 
there are no $O(4)$ scalars in the matter fields in 
adjoint representations.}
This solution can be easily understood by using the standard techniques
\cite{Li}.
Since $\Phi$ is a traceless 
$4\times 4$ hermitean matrix, one can always make an $SU(4)$ gauge 
transformation in order to diagonalize $\Phi$, while preserving
the minimum condition (\ref{minimum}).
There are several non--equivalent SSB patterns
associated with the three independent real eigenvalues 
of the diagonal matrix $Diag(\Phi)$.
In particular, the  solution which leaves unbroken the $O(4)$ 
rotational group, corresponds to the SSB pattern 
$\langle \Phi\rangle \propto Diag(1,1,-1,-1)$, which is proportional to 
the $\gamma_5$ matrix in the so-called chiral basis.

After shifting the scalar field $\bar{\phi}$  around its minimum,
$\bar{\phi}=\langle \bar{\phi}\rangle
+\bar{\eta}$, the spectrum of the theory is manifest.
The gauge fields in Eq.(\ref{Lpure}) which remain massless
are the $\bar{A}_{\mu}$ and  $C_{\mu\alpha\beta}$ ones, since the corresponding
generators (respectively $\gamma_5$ and $\sigma_{\mu\nu}$)
commute with $\gamma_5$, being $\gamma_5$
the generator associated to $\langle \bar{\phi}\rangle$.
Therefore, the $SU(4)$ gauge group is broken to the exact one 
$SO(4)\times U(1)$.
On the contrary, due to the Higgs mechanism, the tensor fields 
$T_{\mu\nu}$ and $\barT_{\mu\nu}$ acquire a mass term, since the
corresponding generators (respectively $\gamma_{\mu}$ and
$\gamma_{\mu}\gamma_5$) do not commute with $\gamma_5$.
Note that this result is in agreement with 
the SSB patterns of $SU(N)$, 
see Ref.\cite{Li} for further details.
Indeed, by means of matter fields in the adjoint representation,
$SU(4)$ can be broken to $SU(2)\times SU(2)\times U(1)$, 
which is isomorphic to $SO(4)\times U(1)$ \cite{Li}.

It is straightforward to generalize the above results to Minkowski space.
By using in this space the same form of the potential (\ref{VEV}), 
the largest 
extended gauge group (which is $SU(2,2)$ \cite{eg})
can be broken to the non--compact one 
$SO(3,1)\times U(1)$, while the Lorentz invariance remains exact.
\section{Conclusions}
In this article we analyzed the free particle spectrum 
of the extended YM gauge theories in Euclidean space.
We found that the physical degrees of freedom contained in the
pure gauge sector of the $SU(4)$ group correspond to the  
spin contents of the following massless 
fields: 3 spin-2, 8 spin-1 and 8 spin-0. Some of the spin-1 and spin-0 fields
correspond to the longitudinal polarizations of the tensor fields 
which cannot be gauged out.
Moreover we analyzed some quantum aspects 
by providing the functional quantization in Euclidean space for a general class
of covariant gauges. 

We presented the results for the $N=1$ and $N=2$ supersymmetric extensions 
of the action in Ref.\cite{eg}.
Moreover we found that the maximum semi-integer spin of the 
supersymmetric extensions is 3/2 (in four dimension)
and these Rarita-Schwinger fields 
could be regarded as the supersymmetric
partners of the spin-2 fields present in the pure gauge theory.

The generalization so as to include the standard internal symmetries 
introduces a larger unified group. In this model the corresponding
algebra of this group is given by the tensorial product of the
Clifford and the internal symmetry algebra.
We give the expression of the extended YM lagrangian generalized so as
to include an $SU(N)$ internal group. 
An interesting consequence of this unified algebra is that,
in addition to the antisymmetric structure functions $f^{abc}$ of 
the internal $SU(N)$ group, 
also the complete symmetric ones $d^{abc}$ appear in the lagrangian
(\ref{LagIS1}-\ref{LagIS2}).

We analyzed the matter fields which transform 
as the fundamental representation
of the extended gauge symmetry group.
In this case we found that these fields can only be of bosonic type.
Moreover it is important to stress
that new kind of gauge--invariant and 
renormalizable couplings can be
generated by means of these matter fields (see Eq.(\ref{Lsigma})),
couplings which have no counterpart in any known gauge theory.
Finally we analyzed the SSB of the extended gauge symmetry.
In particular we show that there exists  a special SSB pattern where the 
extended $SU(4)$ gauge symmetry can be broken to $SO(4)\times U(1)$,
while leaving exact the $O(4)$ rotational invariance.

We conclude the present section by commenting on the
possible future developments of this study.
It would be worth investigating the analytical continuation of this model
in Minkowski space by analysing the unitarity of the $S$ matrix
in perturbation theory. The understanding of this issue could be helpful
in clarifying the relation between unitarity and renormalizability in
the higher spin interactions.
Moreover, other interesting aspects of this model,
which we believe are worth investigating, are
the instanton solutions, and the possible embedding 
of this theory in a string framework.

\section*{Acknowledgments}
I thank E. Alvarez, C. Di Cairano, A. Gonz\'alez-Arroyo, 
L. Ib\'a\~nez, M. Klein, C. Mu\~noz, C. Pena, M. Porrati 
and J.A.M. Vermaseren for useful discussions.
I acknowledge the CERN theory division, where
part of this work was completed, for its kind hospitality.
I acknowledge the financial supports of the TMR network, project
``Physics beyond the standard model'', FMRX-CT96-0090 and the partial
financial support of the CICYT project ref. AEN97-1678.

\newpage
\section*{Appendix}
In this appendix we give the expression of the lagrangian ${\bf L^{IS}}$
in (\ref{LagIS}) 
in terms of the components of the gauge potential (\ref{AIS}) 
along the $Y$ algebra basis.
By using the commutation rules of the $Y$ algebra 
we obtain the following result for ${\bf L^{IS}}$
\footnote{This result, together with some other results in the paper,
have been obtained by means of the algebraic manipulation program Form
\cite{form}.}
\bea
{\bf L^{{\scriptstyle IS}}}&=&\frac{1}{4}\left\{
{\bf F}^{{\bf a}}_{\mu\nu} {\bf F}^{{\bf a}}_{\mu\nu}+
{\bf \barF}^{{\bf a}}_{\mu\nu} {\bf \barF}^{{\bf a}\mu\nu}+
{\bf F}^{{\bf a}}_{\mu\nu\alpha} {\bf F}^{{\bf a}\mu\nu\alpha}+
{\bf \barF}^{{\bf a}}_{\mu\nu\alpha} 
{\bf \barF}^{{\bf a}\mu\nu\alpha}+
{\bf F}^{{\bf a}}_{\mu\nu\alpha\beta} 
{\bf F}^{{\bf a}\mu\nu\alpha\beta}\right\}\no
&+&\frac{1}{4 C_F}\left\{
\bf{\barF}_{\mu\nu} {\bf \barF}^{\mu\nu}+
{\bf F}_{\mu\nu\alpha} {\bf F}^{\mu\nu\alpha}+
{\bf \barF}_{\mu\nu\alpha} {\bf \barF}^{\mu\nu\alpha}+
{\bf F}_{\mu\nu\alpha\beta} {\bf F}^{\mu\nu\alpha\beta}\right\},
\label{LagIS1}
\eea
where $C_F$ is defined in (\ref{Talgebra}) and
the sum over the repeated indices is assumed.
The expressions of the fields strength ${\bf F_i^{a}}$ are given by
\bea
{\bf F}^{{\bf a}}_{\mu\nu}&=&\partial_{\mu} A^{{\bf a}}_{\nu}-\partial_{\nu} 
A^{{\bf a}}_{\mu}-
g f^{\bf abc}\left[A^{{\bf b}}_{\mu}A^{{\bf c}}_{\nu}+
\bar{A}^{{\bf b}}_{\mu}\bar{A}^{{\bf c}}_{\nu}+
T^{{\bf b}}_{\mu\alpha}T^{{\bf c}\alpha}_{\nu}\right.\no
&+&\left.\barT^{{\bf b}}_{\mu\alpha}\barT^{{\bf c}\alpha}_{\nu}+
C^{{\bf b}}_{\mu\alpha\beta}C^{{\bf c}\alpha\beta}_{\nu}
\right],\no
{\bf \barF}_{\mu\nu}&=&\barF_{\mu\nu} - 
g 2 C_F\left[ T^{{\bf a}}_{\mu\alpha}
\barT^{{\bf a}\alpha}_{\nu}-(\mu \leftrightarrow \nu)
\right],\no
{\bf \barF}^{{\bf a}}_{\mu\nu}&=&\partial_{\mu} 
\bar{A}^{{\bf a}}_{\nu}-\partial_{\nu} \bar{A}^{{\bf a}}_{\mu} -
g\left\{ f^{\bf abc}\left[\bar{A}^{{\bf b}}_{\mu}A^{{\bf c}}_{\nu}-
\frac{1}{4}\bar{C}_{\mu\alpha\beta}^{{\bf b}}C_{\nu}^{{\bf c}\alpha\beta}
\right]\right.\no
&+&\left. d^{\bf abc}\left( T^{\bf b}_{\mu\alpha} \barT^{\bf c\alpha}_{\nu}\right)
+2\left[T^{{\bf a}}_{\mu\alpha}\barT^{~\alpha}_{\nu}-
\barT^{{\bf a}}_{\mu\alpha}T^{~\alpha}_{\nu}\right]-\munu
\right\},\no
%
{\bf F}_{\mu\nu\alpha}&=&F_{\mu\nu\alpha}+ g 2 C_F \left[
\bar{A}^{{\bf a}}_{\mu}\barT^{\bf a}_{\nu\alpha}-\sqrt{2}T_{\mu}^{{\bf a}\beta}
C^{{\bf a}}_{\nu\beta\alpha}-\munu \right],\no
{\bf F}^{{\bf a}}_{\mu\nu\alpha}&=&\partial_{\mu} T^{{\bf a}}_{\nu\alpha}-
\partial_{\nu} T^{{\bf a}}_{\mu\alpha}
-g\left\{ f^{\bf abc}\left[
T_{\mu\alpha}^{{\bf b}}A^{{\bf c}}_{\nu}-
\frac{1}{\sqrt{2}}\barT_{\mu}^{{\bf b}\beta}
\overline{C}_{\nu\alpha\beta}^{{\bf c}}\right]\right.\no
&+&\left.d^{\bf abc}\left[
\barT_{\mu\alpha}^{{\bf b}}\bar{A}^{{\bf c}}_{\nu}-
\sqrt{2}T_{\mu}^{{\bf b}\beta}
C_{\nu\alpha\beta}^{{\bf c}}\right]+2\left[
\barT_{\mu\alpha}\bar{A}^{{\bf a}}_{\nu}+
\barT_{\mu\alpha}^{{\bf a}}\bar{A}_{\nu}\right.\right.\no
&-&\left.\left.\sqrt{2}\left(C_{\nu\alpha\beta}T_{\mu}^{{\bf a}\beta}+
C_{\nu\alpha\beta}^{{\bf a}}T_{\mu}^{~\beta}\right)\right]-\munu
\right\},\no
{\bf\barF}_{\mu\nu\alpha}&=&\barF_{\mu\nu\alpha} - g 2 C_F\left[
\bar{A}^{{\bf a}}_{\mu}T^{\bf a}_{\nu\alpha}+\sqrt{2}\barT_{\mu}^{{\bf a}\beta}
C^{{\bf a}}_{\nu\beta\alpha}-\munu \right],\no
{\bf \barF}^{{\bf a}}_{\mu\nu\alpha}&=&
\partial_{\mu} \barT^{{\bf a}}_{\nu\alpha}-
\partial_{\nu} \barT^{{\bf a}}_{\mu\alpha}
-g\left\{ f^{\bf abc}\left[
\barT_{\mu\alpha}^{{\bf b}}A^{{\bf c}}_{\nu}+
\frac{1}{\sqrt{2}}T_{\mu}^{{\bf b}\beta}
\bar{C}_{\nu\alpha\beta}^{{\bf c}}\right]\right.\no
&+&\left.d^{\bf abc}\left[
T_{\nu\alpha}^{{\bf b}}\bar{A}^{{\bf c}}_{\mu}-
\sqrt{2}\barT_{\mu}^{{\bf b\beta}}C_{\nu\alpha\beta}^{{\bf c}}\right]+
2\left[T_{\nu\alpha}\bar{A}^{{\bf a}}_{\mu}-
T_{\mu\alpha}^{{\bf a}}\bar{A}_{\nu}\right.\right.\no
&-&\left.\left.\sqrt{2}\left(C_{\nu\alpha\beta}\barT_{\mu}^{{\bf a}\beta}-
C_{\mu\alpha\beta}^{{\bf a}}\barT_{\nu}^{~\beta}\right)\right]-\munu
\right\},\no
%
{\bf F}_{\mu\nu\alpha\beta}&=&F_{\mu\nu\alpha\beta} +
g \sqrt{2} C_F\left(T^{{\bf a}}_{\mu\alpha}T^{{\bf a}}_{\nu\beta}+
\barT^{{\bf a}}_{\mu\alpha}\barT^{{\bf a}}_{\nu\beta}-
2 C_{\mu\gamma\alpha}^{{\bf a}}C^{{\bf a}~\gamma}_{\nu\beta}-
\munu \right),\no
{\bf F}^{{\bf a}}_{\mu\nu\alpha\beta}&=&\partial_{\mu} 
C^{{\bf a}}_{\nu\alpha\beta}-
\partial_{\nu} C^{{\bf a}}_{\mu\alpha\beta}
+ g\left\{ f^{\bf abc}\left[\frac{1}{2} \bar{A}_{\mu}^{{\bf b}}
\bar{C}_{\nu\alpha\beta}^{{\bf c}}+\frac{1}{\sqrt{2}}T^{{\bf b}}_{\mu\gamma}
\barT^{{\bf c}}_{\nu\delta}
\epsilon_{\alpha\beta}^{~~~\gamma\delta}+C_{\mu\beta\alpha}^{{\bf b}}
A^{{\bf c}}_{\nu}\right.\right.\no
&-&\left.\left. \munu\right]
+ d^{\bf abc}\frac{1}{\sqrt{2}}\left[
T^{{\bf b}}_{\mu\alpha}T^{{\bf c}}_{\nu\beta}+
\barT^{{\bf b}}_{\mu\alpha}\barT^{{\bf c}}_{\nu\beta}-
2 C_{\mu\gamma\alpha}^{{\bf b}}C^{{\bf c}~\gamma}_{\nu\beta}-
\albe\right]\right.\no
&+&\left.\sqrt{2}\left[\left(T^{{\bf a}}_{\mu\alpha}T_{\nu\beta}+
\barT^{{\bf a}}_{\mu\alpha}\barT_{\nu\beta}-
2C_{\mu\gamma\alpha}^{{\bf a}} C_{\nu\beta}^{~~\gamma}-\munu\right)
-\albe \right]
\right\},
\label{LagIS2}
\eea
where the expressions for 
$\barF_{\mu\nu}$, $F_{\mu\nu\alpha}$, $\barF_{\mu\nu\alpha}$,
and $F_{\mu\nu\alpha\beta}$ are given in (\ref{fieldtrengths})
and $\bar{C}_{\mu\alpha\beta}\equiv\epsilon_{\alpha\beta\gamma\delta} 
C_{\mu}^{~\gamma\delta},~~
\bar{C}_{\mu\alpha\beta}^{\bf a}\equiv
\epsilon_{\alpha\beta\gamma\delta} 
C_{\mu}^{{\bf a}\gamma\delta}$.
The lagrangian ${\bf L^{IS}}$ is invariant under the following 
infinitesimal gauge transformations
\bea
{\bf \delta A_{\mu}^{{\bf a}}}&=&-\frac{1}{g}
\partial_{\mu}\epsilon^{{\bf a}} + f^{\bf abc}\left[
\bar{A}_{\mu}^{{\bf b}}\bar{\epsilon}^{{\bf c}}+
A_{\mu}^{{\bf b}}\epsilon^{{\bf c}}+
T_{\mu\alpha}^{{\bf b}}\epsilon^{{\bf c}\alpha}+
\barT_{\mu\alpha}^{{\bf b}}\bar{\epsilon}^{{\bf c}\alpha}+
C_{\mu\alpha\beta}^{{\bf b}}\epsilon^{{\bf c}\alpha\beta}\right],\no
{\bf \delta \bar{A}_{\mu}}&=&\delta \bar{A}_{\mu} - 2C_F\left[
-T^{{\bf a}}_{\mu\alpha}\bar{\epsilon}^{{\bf a}\alpha}+
\barT^{{\bf a}}_{\mu\alpha}\epsilon^{{\bf a}\alpha}\right],\no
{\bf \delta \bar{A}_{\mu}^{{\bf a}}}&=&-\frac{1}{g}
\partial_{\mu}\epsilon^{{\bf a}} - \left\{f^{\bf abc}\left[
-\bar{A}_{\mu}^{{\bf b}}\epsilon^{{\bf c}}-
A_{\mu}^{{\bf b}}\bar{\epsilon}^{{\bf c}}+
\bar{C}_{\mu\alpha\beta}^{{\bf b}}\epsilon^{{\bf c}\alpha\beta}\right]\right.
\no
&+&\left.d^{\bf abc}\left[-T_{\mu\alpha}^{{\bf b}}\bar{\epsilon}^{{\bf c}\alpha}+
\barT_{\mu\alpha}^{{\bf b}}\epsilon^{{\bf c}\alpha}\right]+2\left[
-T_{\mu\alpha}^{{\bf a}}\bar{\epsilon}^{\alpha}+
\barT_{\mu\alpha}^{{\bf a}}\epsilon^{\alpha}
-T_{\mu\alpha}\bar{\epsilon}^{{\bf a}\alpha}+
\barT_{\mu\alpha}\epsilon^{{\bf a}\alpha}\right]\right\},\no
{\bf \delta T_{\mu\alpha}}&=&\delta T_{\mu\alpha} - 2C_F\left[
\bar{A}_{\mu}^{{\bf a}}\bar{\epsilon}^{{\bf a}}_{\alpha}-
\barT^{{\bf a}}_{\mu\alpha}\bar{\epsilon}^{{\bf a}}+
\sqrt{2}\left(T^{{\bf a}}_{\mu\beta}\epsilon^{{\bf a}\beta}_{\alpha}
+C_{\mu\gamma\alpha}^{{\bf a}}\epsilon^{{\bf a}\gamma}\right)\right],\no
{\bf \delta T^{{\bf a}}_{\mu\alpha}}&=&-\frac{1}{g}
\partial_{\mu}\epsilon^{{\bf a}}_{\alpha} - 
\left\{f^{\bf abc}\left[-T_{\mu\alpha}^{{\bf b}}\epsilon^{{\bf c}}-
A_{\mu}^{{\bf b}}\epsilon^{{\bf c}}_{\alpha}+
\frac{1}{\sqrt{2}}\left(
\bar{C}_{\mu\alpha\delta}^{{\bf b}}\bar{\epsilon}^{{\bf c}\delta}
+T^{{\bf b}}_{\mu\delta}
\bar{\epsilon}_{\alpha}^{{\bf c}\delta}\right)\right]\right.\no
&+&\left.d^{\bf abc}\left[\bar{A}^{{\bf b}}_{\mu}\bar{\epsilon}^{{\bf c}}_{\alpha}-
\barT_{\mu\alpha}^{{\bf b}}\bar{\epsilon}^{{\bf c}}+
\sqrt{2}\left(T^{{\bf b}}_{\mu\delta}\epsilon^{{\bf c}\delta}_{\alpha}+
C_{\mu\delta\alpha}^{{\bf b}}\epsilon^{{\bf c}\delta}\right)\right]\right.\no
&+&\left.2\left[\bar{A}_{\mu}^{{\bf a}}\bar{\epsilon}_{\alpha}+
\bar{A}_{\mu}\bar{\epsilon}_{\alpha}^{{\bf a}}-
\barT_{\mu\alpha}^{{\bf a}}\bar{\epsilon}-\barT_{\mu\alpha}\bar{\epsilon}^{{\bf a}}
\right.\right.\no
&+&\left.\left.\sqrt{2}\left(T^{{\bf a}}_{\mu\delta}\epsilon_{\alpha}^{~\delta}+
T_{\mu\delta}\epsilon^{{\bf a}\delta}_{\alpha}+
C_{\mu\delta\alpha}^{{\bf a}}\epsilon^{\delta}+
C_{\mu\delta\alpha}\epsilon^{{\bf a}\delta}\right)\right]
\right\},\no
{\bf \delta \barT_{\mu\alpha}}&=&\delta \barT_{\mu\alpha} - 2C_F\left[
-\bar{A}_{\mu}^{{\bf a}}\epsilon^{{\bf a}}_{\alpha}
+T^{{\bf a}}_{\mu\alpha}\bar{\epsilon}^{{\bf a}}+
\sqrt{2}\left(\barT^{{\bf a}}_{\mu\beta}\epsilon^{{\bf a}\beta}_{\alpha}+
C_{\mu\gamma\alpha}^{{\bf a}}\bar{\epsilon}^{{\bf a}\gamma}\right)\right],\no
{\bf \delta \barT^{{\bf a}}_{\mu\alpha}}&=&-\frac{1}{g}
\partial_{\mu}\bar{\epsilon}^{{\bf a}}_{\alpha}
- \left\{f^{\bf abc}\left[-\barT_{\mu\alpha}^{{\bf b}}\epsilon^{{\bf c}}-
A_{\mu}^{{\bf b}}\bar{\epsilon}^{{\bf c}}_{\alpha}
-\frac{1}{\sqrt{2}}\left(
\bar{C}_{\mu\alpha\delta}^{{\bf b}}\epsilon^{{\bf c}\delta}
+T^{{\bf b}}_{\mu\delta}\bar{\epsilon}_{\alpha}^{{\bf c}\delta}\right)\right]\right.\no
&+&\left.d^{\bf abc}\left[-\bar{A}^{{\bf b}}_{\mu}\epsilon^{{\bf c}}_{\alpha}+
T_{\mu\alpha}^{{\bf b}}\bar{\epsilon}^{{\bf c}}+
\sqrt{2}\left(\barT^{{\bf b}}_{\mu\delta}\epsilon^{{\bf c}\delta}_{\alpha}+
C_{\mu\delta\alpha}^{{\bf b}}\bar{\epsilon}^{{\bf c}\delta}\right)\right]\right.\no
&+&\left.2\left[-\bar{A}_{\mu}^{{\bf a}}\epsilon_{\alpha}-
\bar{A}_{\mu}\epsilon_{\alpha}^{{\bf a}}+
T_{\mu\alpha}^{{\bf a}}\bar{\epsilon}+
T_{\mu\alpha}\bar{\epsilon}^{{\bf a}}
\right.\right.\no
&+&\left.\left.\sqrt{2}\left(
\barT^{{\bf a}}_{\mu\delta}\epsilon_{\alpha}^{~\delta}+
\barT_{\mu\delta}\epsilon^{{\bf a}\delta}_{\alpha}+
C_{\mu\delta\alpha}^{{\bf a}}\bar{\epsilon}^{\delta}+
C_{\mu\delta\alpha}\bar{\epsilon}^{{\bf a}\delta}\right)\right]
\right\},\no
{\bf \delta C_{\mu\alpha\beta}}&=&\delta C_{\mu\alpha\beta} -
2C_F\left\{\frac{1}{\sqrt{2}}\left(T^{{\bf a}}_{\mu\alpha}
\epsilon^{{\bf a}}_{\beta}
+\barT^{{\bf a}}_{\mu\alpha}\bar{\epsilon}^{{\bf a}}_{\beta}\right)-
\sqrt{2}C^{{\bf a}}_{\mu\delta\alpha}\epsilon^{{\bf a}\delta}_{\beta}-\albe
\right\},\no
{\bf \delta C^{{\bf a}}_{\mu\alpha\beta}}&=&-\frac{1}{g}
\partial_{\mu}\epsilon^{{\bf a}}_{\alpha\beta} - \left\{
f^{\bf abc}\left[
C_{\mu\beta\alpha}^{{\bf b}}\epsilon^{{\bf c}}+\frac{1}{2}\left(
-\bar{C}^{\bf b}_{\mu\beta\alpha}\bar{\epsilon}^{{\bf c}}+
\bar{A}_{\mu}^{{\bf b}}\bar{\epsilon}^{{\bf c}}_{\alpha\beta}\right)
-A_{\mu}^{\bf b} \epsilon^{\bf c}_{\alpha\beta}
\right.\right.\no
&+&\left.\left.\frac{1}{\sqrt{2}}\left(
T^{{\bf b}}_{\mu\delta}\bar{\epsilon}^{{\bf c}}_{\gamma}-
\barT^{{\bf b}}_{\mu\delta}\epsilon^{{\bf c}}_{\gamma}\right)
\epsilon_{\alpha\beta}^{~~~\delta\gamma}
\right]\right.\no
&+&\left.d^{\bf abc}\left[\frac{1}{\sqrt{2}}
\left(T^{{\bf b}}_{\mu\alpha}\epsilon^{{\bf c}}_{\beta}+
\barT^{{\bf b}}_{\mu\alpha}\bar{\epsilon}^{{\bf c}}_{\beta}\right)+
\sqrt{2}C_{\mu\alpha\delta}^{{\bf b}}\epsilon^{{\bf c}
\delta}_{\beta}-\albe\right]
\right.\no
&+&\left.\sqrt{2}\left[T^{{\bf a}}_{\mu\alpha}\epsilon_{\beta}+
\barT^{{\bf a}}_{\mu\alpha}\bar{\epsilon}_{\beta}+
T_{\mu\alpha}\epsilon^{{\bf a}}_{\beta}+
\barT_{\mu\alpha}\bar{\epsilon}^{{\bf a}}_{\beta}+
2\left(C_{\mu\alpha\delta}^{{\bf a}}\epsilon^{~\delta}_{\beta}+
C_{\mu\alpha\delta}\epsilon^{{\bf a}\delta}_{\beta}\right)
\right.\right.\no
&-&\left.\left.\albe \right]
\right\},
\eea
where the expressions for $\delta \bar{A}_{\mu},~
\delta T_{\mu\nu},~\delta \barT_{\mu\nu}$, and $\delta C_{\mu\alpha\beta}$
are given in (\ref{gaugetransf}), and
$\bar{\epsilon}_{\alpha\beta}\equiv \epsilon_{\alpha\beta\gamma\delta}
\epsilon^{\gamma\delta}$,
$\bar{\epsilon}^{\bf a}_{\alpha\beta}\equiv 
\epsilon_{\alpha\beta\gamma\delta}\epsilon^{{\bf a}\gamma\delta}$.
(To avoid confusion we recall that $\epsilon_{\alpha\beta\gamma\delta}$
is the complete antisymmetric tensor in Euclidean space.)
In the lagrangian (\ref{LagIS1}) (and therefore in the Feynman rules)
the contractions of the type
$d^{\bf abc} f^{\bf cef}$, which cannot be reduced as the product of 
combinations of $\delta^{\bf ab},~d^{\bf abc}$, and $f^{\bf abc}$
for a  general $SU(N)$ group \cite{vprivate}, appear.
(Note that simplifications can be obtained by choosing 
some particular internal groups).
Clearly the physical amplitudes and correlation functions can be expressed as
products of $\delta^{\bf ab}$. Useful results and techniques are 
provided in Refs.\cite{colours} for the calculations of the 
traces of the  ${\bf T^a}$ matrix products.
\end{document}